\documentclass[10pt,letterpaper]{IEEEtran}
\usepackage{amssymb,amsmath}
\usepackage{color,soul}
\usepackage{amsfonts}
\usepackage{array}
\usepackage{algorithmicx}
\usepackage[ruled,commentsnumbered]{algorithm2e}
\usepackage[nodisplayskipstretch]{setspace}
\usepackage{mathrsfs}
\usepackage{graphicx}
\usepackage{multirow}
\usepackage{epstopdf}
\usepackage{epsfig}
\usepackage{stmaryrd}
\usepackage{multirow,url}
\usepackage{color,cite}
\usepackage{amsthm}
\usepackage{bm}
\usepackage{pgf,tikz}
\usepackage{subfigure,cite}

\theoremstyle{definition}

\theoremstyle{remark}

\usepackage{fancyhdr,ifthen}

\DeclareMathOperator{\Exp}{\mathbb{E}}

\newcounter{section:outage-analysis}
\begin{document}

\title{Multi-Source Cooperative Communication with Opportunistic Interference Cancelling Relays \thanks{The author is with the Department of Electrical and Computer Engineering, University of Thessaly, 37 Glavani - 28th October Street
T.K. 38221, Volos, Greece, Greece (Email: anargyr@ieee.org).}
}

\author{Antonios Argyriou, \textit{Senior Member, IEEE} \vspace{-6mm}}

\maketitle
\graphicspath{{figures/}}
\markboth{\today, to appear in IEEE Transactions on Communications}{\today, to appear in IEEE Transactions on Communications}

\begin{abstract}
In this paper we present a multi-user cooperative protocol for wireless networks. Two sources transmit simultaneously their information blocks and relays employ opportunistically successive interference cancellation (SIC) in an effort to decode them. An adaptive decode/amplify-and-forward scheme is applied at the relays to the decoded blocks or their sufficient statistic if decoding fails. The main feature of the protocol is that SIC is exploited in a network since more opportunities arise for each block to be decoded as the number of used relays $N_\text{RU}$ is increased. This feature leads to benefits in terms of diversity and multiplexing gains that are proven with the help of an analytical outage model and a diversity-multiplexing tradeoff (DMT) analysis. The performance improvements are achieved without any network synchronization and coordination. In the final part of this work the closed-form outage probability model is used by a novel approach for offline pre-selection of the $N_\text{RU}$ relays, that have the best SIC performance, from a larger number of $N_\text{R}$ nodes. The analytical results are corroborated with extensive simulations, while the protocol is compared with orthogonal and multi-user protocols reported in the literature.
\end{abstract}


\section{Introduction}
One recurring research theme in the area of wireless networks has been the optimal use of every node for the benefit of the complete network. The cooperative communication paradigm is one such approach, where relay nodes forward signals between other nodes to improve their performance~\cite{meulen68,sendonaris98}. The main benefit is a \emph{diversity gain}~\cite{sendonaris98} that is essential for improving performance in slow fading channels. When multiple relays are available, they can create a \emph{distributed antenna system} to mimic the behavior of the classic multi-antenna MIMO systems that accomplish the same with space-time coding (STC)~\cite{alamouti98,tarokh99}. In this case the relays receive a signal from a source, and then they can construct a distributed space-time code (DSTC)~\cite{laneman03,hua03,chang04,jing06}. The first works on DSTC considered relays that decode the received signals and then create the DSTC based on this result~\cite{laneman03,hua03,chang04}. Besides this decode-and-forward (DF) DSTC strategy (referred as DSTC/DF), more general DSTCs for amplify-and-forward (AF) relays were introduced in~\cite{jing06}. The work in~\cite{jing06} allows a single transmitter to broadcast an information block in one slot, while in the subsequent slot the relays encode their received signals into a distributed linear dispersion (LD) code. One key characteristic of this work is that the relays do not decode the received signals, hence the name DSTC/AF. Overall DSTC leads to a higher receiver SNR (diversity gain) but all forms of STC include also a multiplexing gain. Their relative interplay is captured with the diversity-multiplexing tradeoff (DMT)~\cite{book:fundamental-wireless}. Higher spectral efficiency (SE) is possible since the higher SNR can be combined with higher order modulation and coding schemes (MCSs). However, in a two-hop network, the multiplexing gain that can be achieved with one source is equal to $\frac{1}{2}$~\cite{laneman03}. Hence, the classic relay-based systems are limited in terms of the potential multiplexing gain.

\begin{figure}[t]
\centering
\subfigure[]{\includegraphics[keepaspectratio,width = 0.45\linewidth]{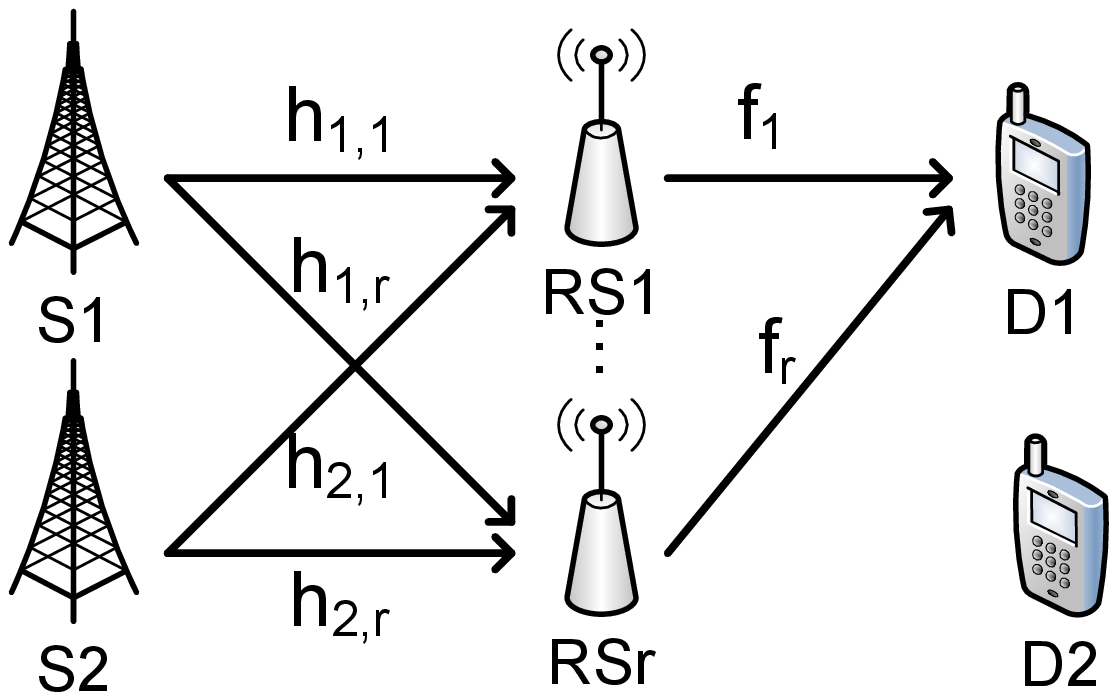}\label{fig:main-idea}}
\subfigure[]{\includegraphics[keepaspectratio,width = 0.45\linewidth]{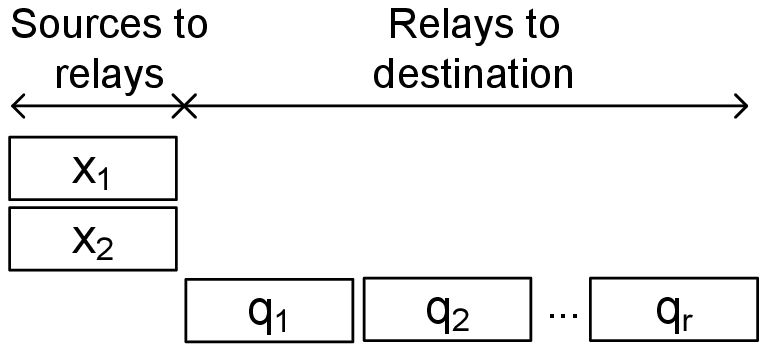}\label{fig:protocol}}
\caption{a) The network model in this paper consists of a multiple access multi-relay network (MAMRN). In this figure the MAMRN is mapped to a downlink transmission scenario. b) Cooperative protocol transmissions in the time domain. }
\end{figure}

The natural question is if one can exploit nodes that act as relays in a \emph{multi-user network} with non-orthogonal transmission protocols that can lead to higher multiplexing gains. The relay network model in this case consists of a cascaded compound multiple access channel (C-MAC) in the first hop (the message from each source needs to be decoded at every relay~\cite{ahlswede74}), and a MAC in the second hop (Fig.~\ref{fig:main-idea}). This is also referred to as a multiple access multi-relay network (MAMRN). This setup has many applications in cellular in-band relay-based uplink and downlink (Fig.~\ref{fig:main-idea})~\cite{yang10}, relay-based downlink multicast communication~\cite{jnl_2010_twc} (also in Fig.~\ref{fig:main-idea} with the addition of D2), and cellular device-to-device (D2D) communication where the relays can be users themselves~\cite{LTE12}.

Multi-source communication in this MAMRN can be handled efficiently with \emph{network MIMO} techniques~\cite{gesbert10}. DF and AF relay protocols for uplink cellular communication were analyzed in~\cite{simeone08,gesbert10,letaief12}. In~\cite{simeone08} it was shown that DF relays for the MAMRN perform very well in terms of outage for slow fading channels in the low SNR regime, while AF performs better in the high SNR regime due to the lack of noise amplification. The work in~\cite{simeone08} is essentially the implementation of Han and Kobayashi DF scheme but for a cellular MAMRN. A third compress-and-forward (CF) scheme, originally proposed in~\cite{gastpar04,gastpar05}, was shown to be better than AF and DF for the MAMRN in~\cite{simeone10}, under the assumptions of a non-fading channel, full-duplex relay, and multicell processing (in our scenario two destinations would be needed to communicate through a backhaul link). Hence, this CF scheme cannot be implemented directly in our setup since it requires side-information from a second BS. Another important result in~\cite{simeone10}, was that when the relay power is the same as the source, then CF performs slightly worse than AF even under multicell processing.  More recently, the authors in~\cite{wei12} considered a MAMRN with the use of structured lattice codes at the sources and AWGN links on the second hop.

Although the topology we consider uses two hops, several works have focused on multi-user communication in single hop networks usually with the presence of a single relay, but also without one. The first scenario is typically referred to as the multiple access relay channel (MARC). Some of these protocols we discuss below can be adapted to operate in two-hop networks. These protocols are similarly categorized as DF/AF. Overall in a MARC higher multiplexing gains can be achieved as shown in~\cite{belfiore07,badr08,yao12,giannakis08,Wei15}. In~\cite{giannakis08} the authors proposed complex field network coding (CFNC) for the MARC. The relay uses DF and symbol-level network coding over a complex field for the symbols originating from the two sources. This protocol requires the pre-distribution of the coefficients in the complete network. Furthermore, in the same work extensions for multiple relays and sources are provided while the protocol is inherently engineered to support a single destination. Another recent work reported in~\cite{yao12}, proposed a DF-based DSTC protocol for a single hop network that consisted of two sources and one receiver without the presence of a relay. The protocol achieves a multiplexing gain equal to $\frac{2}{3}$ for a half-duplex system, and 1 for a full-duplex system. This last work will serve as an interesting benchmark for our system. Regarding AF-based protocols, the authors in~\cite{jnl_2010_twc} proposed an analog network coding (ANC) based protocol for the \emph{multicast MARC} channel. For the MARC with a single destination, the authors in~\cite{Wei15} generalized the concept of ANC with relays that execute ANC in a complex field (i.e., similar to CFNC). A transformation operation is executed at the relay based on feedback received from the destination. In the case of a multi-relay network the extension is not directly obvious (hence difficult to collect full diversity), while the increased rate of transmitter channel state information (CSIT) needed at the relays, might be a problem similar to~\cite{giannakis08}. In~\cite{belfiore07,badr08} AF-based DSTCs with a special non-orthogonal structure for the two-user MARC were proposed.

In this paper we propose a multi-user cooperative protocol for the MAMRN in Fig.~\ref{fig:main-idea}. We allow the two sources to transmit simultaneously without any coordination with the remaining network (e.g., precoding, beamforming), or precise symbol/bit/packet-level synchronization. Interfering signals are exploited in the MAMRN by attempting to decode at each relay both simultaneously-transmitted information blocks. This is accomplished by using the optimal decoding strategy at each relay for this case which is successive interference cancellation (SIC)~\cite{tse04}. Next, we apply an adaptive DF/AF protocol both on decoded blocks but also on the non-decoded blocks. The sources and relays do not require any form of CSIT while the communication model in the second hop can be either unicast or multicast~\cite{jnl_2010_twc}. We model rigorously the performance the proposed protocol that its preliminary version was first reported in~\cite{cnf_2014_globalsip}. Our transmission protocol is engineered with a mindset towards simplicity and full exploitation of potential relays without performance compromise.

\textbf{Contributions/Main results.} The contributions of this work are: 1) \emph{A novel multi-user protocol design for relay networks}. In the high SNR regime our DMT analysis, corroborated with simulation, reveals that \emph{our protocol achieves a multiplexing gain of $\frac{2}{3}$ that is typically possible with one-hop DSTC systems}~\cite{giannakis08,yao12}. The above is true even in strong interference conditions between the two sources. In the low SNR regime our protocol outperforms classic orthogonal DSTCs and multi-user protocols. 2) \emph{ A rigorous outage analysis for the proposed protocol} that jointly considers interfering transmissions, SIC decoding, and the AF/DF relay operations. The performance analysis constitutes on its own a contribution of this paper since \emph{it characterises the performance of multirate transmitters and SIC decoding}. The overall model is used first for shedding light into the inner workings of our protocol, and second for the DMT analysis, while it is validated with simulations. 3) We also work towards extending our idea in more general networks with multiple relays. In particular we propose \emph{a relay pre-selection technique} that uses our outage model for selecting nodes statically as suitable relay candidates based on their ability to decode with SIC. The ramifications of the last idea are important since it provides a path for a power-efficient and low-complexity use of nodes in dense wireless networks.

\textbf{Paper Organization.} The rest of the paper is organized as follows. The system model and a detailed description of the proposed cooperative protocol can be found in Section~\ref{section:protocol}. The outage performance analysis for the protocol is presented in Section~\ref{section:outage-analysis}, while the DMT analysis is in Section~\ref{section:dmt}. The use of our model for optimization in relay networks is presented in Section~\ref{section:network}. Section~\ref{section:performance-evaluation} presents simulation and numerical results, while Section~\ref{section:conclusions} presents our conclusions.

\section{Proposed Cooperative System}
\label{section:protocol}
\subsection{System Model}
We consider a relay network model, depicted in Fig.~\ref{fig:main-idea}, with two sources S1 and S2, and $N_\text{R}$ relays denoted as RS1, RS2, ...., RSr. We examine a many-to-one communication scenario (S1,S2$\rightarrow$D1) in this paper. However, our protocol and our analysis are applicable, and without any modification, to a unicast scenario (S1$\rightarrow$ D1,S2$\rightarrow$ D2), and also multicast (S1,S2$\rightarrow$ D1,D2,...). All nodes are equipped with a single omni-directional antenna that can be used both for transmission and reception in half duplex mode, while all nodes have the same average power constraint. We denote the independent channels from the $s$-th source to the $r$-th relay as $h_{s,r}$, and the channel from the $r$-th relay to destination as $f_{r}$. These channels are block-fading (quasi-stationary) and Rayleigh, i.e., $h_{s,r}$$\sim$$\mathcal{CN}(1,1)$, $f_{r}$$\sim$$\mathcal{CN}(1,1)$ (complex Gaussian random variables with unit mean and variance). AWGN with zero mean and variance $\sigma^2$ is assumed at all receivers. Receiver CSI is needed for SIC decoding and for calculating the power scaling at the relays. No further channel knowledge is required. CSI at the final destination is similarly assumed to be available from  training signals or packet preambles~\cite{jing06,laneman03,Jafarkhani11}.

Regarding the paper notation matrices are denoted with bold capital letters while bold lowercase letters denote vectors. Also $\Exp [\cdot]$ is the expectation of a random variable. 


\subsection{Cooperative Protocol}
Since we assume that there is no direct link between the sources and the destination, three time slots are required for the transmission of two blocks (Fig.~\ref{fig:protocol}). In the first slot the sources broadcast simultaneously their blocks denoted as $x_1$, and $x_2$ respectively. The communication rates of the sources are $R_1$ and $R_2$ bits/symbol respectively, and they also take into account both modulation and coding~\cite{tse04}. The power dedicated to each transmitted block from the sources is normalized to unity, i.e., $\Exp[|x_1|^2]$=$\Exp[|x_2|^2]$=1. Thus, the baseband model for the interfering blocks at relay $r$ is $y_r=h_{1,r}x_1+h_{2,r}x_2+w_r$, where $w_r$ is the AWGN sample at relay $r$.

After the first broadcast phase of the protocol is completed, each relay decodes the two blocks by employing ordered SIC. That is, the block with the highest energy/bit is decoded first while the other block is treated as noise while no precise synchronization is needed~\cite{book:fundamental-wireless}. To understand why the energy/bit is used for selecting the decoding order consider that there is no interference. Then, the probability of symbol error for MQAM can be approximated by $ 4Q \Big ( \sqrt{\frac{3\Exp[|x_1|^2]|h_{1,r}|^2}{\sigma^2(2^{R_1}-1)}} \Big ) $~\cite{goldsmith:wireless-communications}. The fractional term inside the Q function is essentially the normalized SNR/bit. We can get a similar expression for the second source. Since we assumed $\Exp[|x_1|^2]$=$\Exp[|x_2|^2]$=1, the following condition is checked so that $x_1$ is decoded first:\footnote{It is possible that different rules are used for selecting the symbol to be decoded first or even a completely different IC scheme can be adopted.}
\begin{equation}
\label{eqn:sic-condition}
\frac{|h_{1,r}|^2}{2^{R_1}-1} > \frac{|h_{2,r}|^2}{2^{R_2}-1}
\end{equation}
If $x_1$ is correctly decoded, it is subtracted from the aggregate signal $y_r$. Regarding the implementation of the cancellation mechanism it is executed at the block level. Thus, upon the successful decoding, and with CSI at the relay (in this example $h_{1,r}$), we can completely remove/cancel a complete block from the aggregate received signal $y_r$.

Depending on the result, the $r$-th relay will transmit different signals in the $r$-th time slot of the protocol (Fig.~\ref{fig:protocol}). To denote these signals that the relays transmit we use the notation $q_{r}$. This relay pre-processing can be compactly modeled as:
\begin{align}\label{eqn:q_ri}
q_{r} &=a_{r,1}x_1+ a_{r,2}x_2 +a_{r,3}w_r
\end{align}
The adopted signal notation covers every possible block decoding outcome at the relay through the complex parameters $a_{r,1},a_{r,2},a_{r,3}$ defined in Table~\ref{table:preprocessing}. In case both blocks are decoded at both relays, then $q_r$ effectively contains the superposition of the blocks $x_1,x_2$.
\begin{table}[t]
\footnotesize
\centering
\begin{tabular}{ |l|l|l|l|l|l|l| }
\hline
decoded?& $x_1,x_2$     & $x_1$                             & $x_2$                             &none \\ %
\hline
$a_{r,1}$ & 1       & 1                                 & $h_{1,r}$                                & $h_{1,r}$\\%
$a_{r,2}$ & 1       & $h_{2,r}$                          & 1                                 & $h_{2,r}$\\
$a_{r,3}$ & 0       & 1  & 1                                 & 1 \\
\hline
\hline
\end{tabular}
\caption{Signal Modeling at the Relays.}
\label{table:preprocessing}
\end{table}
Based on the previous discussion we see that our protocol applies an adaptive AF/DF protocol for signal $q_{r}$ depending on its content (decoded signal or not). In matrix notation the transmitted signal from all the relays is:
\begin{equation*}\label{eqn:Z}
\mathbf{Z}=\text{diag} ( g_{1}q_{1} ~~ g_{2}q_{2}~~...~~ g_{r}q_{r} ~~...)
\end{equation*}
\normalsize
In the above matrix the rows indicate the relay and the columns the time slot, while $g^2_{r}$ is the transmit power for signal $q_{r}$. Now after scaling is applied, the relays broadcast their blocks in $\mathbf{Z}$ sequentially. A total of $N_\text{F}=N_\text{R}+1$ slots are needed.

To define the system at the receiver let us denote with $\mathbf{f}$ the vector of the channel gains for the links between the relays and the destination. The received signal at the destination over $N_\text{R}$ consecutive slots will be the vector:

\small
\begin{align*}
\mathbf{\tilde{y}} 
   &= \left [ \begin{array}{cc}  f_{1}g_{1}q_{1}\\ f_{2}g_{2}q_{2}\\ ...\end{array}\right]^T
  +\left [ \begin{array}{cc}  w(1)\\ w(2)\\...\end{array}\right]^T\nonumber
  \end{align*}
  \normalsize
In the above $w(r)$ are the AWGN samples during the specific slot. Elaborating on the above leads to:
\small
  \begin{align}
  \mathbf{\tilde{y}}
   & =\left [ \begin{array}{cc}
  g_{1}f_1a_{1,1} & g_{2}f_1a_{1,2}  \\
  g_{2}f_2a_{2,1}  & g_{1}f_2 a_{2,2}\\
  ... & ...
  \end{array}\right]\left [ \begin{array}{cc}
  x_1  \\
  x_2
  \end{array}\right]+\left [ \begin{array}{cc}
  g_{1}f_1a_{1,3}w_{r1} \\
  g_{2}f_2a_{2,3}w_{r2}\\
  ...
  \end{array}\right]+\left [ \begin{array}{cc}
   w(1)  \\
  w(2)\\
  ...
  \end{array}\right]\nonumber\\
  &=\mathbf{H}\mathbf{x}+\mathbf{w}
\label{eqn:signal_model1}
\end{align}
\normalsize
The destination still receives $N_\text{R}$ observations over $N_\text{R}$ slots that it can optimally solve with our linear MIMO MMSE-SIC decoder~\cite{tse04} described next.

For decoding based on the signal model in~\eqref{eqn:signal_model1} we calculate the covariance matrix $\mathbf{\Sigma}_{\mathbf{w}}$ of the noise vector. The entries of this $N_\text{R}\times N_\text{R}$ matrix are $[\mathbf{\Sigma}_{\mathbf{w}}]_{r,r'}=0,\forall r\neq r'$ and
$[\mathbf{\Sigma}_{\mathbf{w}}]_{r,r}=g^2_{r}|f_r|^2|a_{r,3}|^2\sigma^2+\sigma^2$. For final decoding of the transmitted symbols we apply linear MMSE equalization for the signal model in~\eqref{eqn:signal_model1} as follows:
\begin{align}
\hat{\mathbf{x}}= \text{HDD} \{ (\mathbf{H}^H\mathbf{\Sigma}^{-1}_{\mathbf{w}}\mathbf{H}+\mathbf{I})^{-1}\mathbf{H}^H\mathbf{\Sigma}^{-1}_{\mathbf{w}}\mathbf{\tilde{y}} \}
\end{align}
In the above HDD stands for hard decision decoding.\footnote{The adoption of HDD intends to lower the execution time of decoding at the relays and the destination and accelerate simulation. Joint demodulation and decoding of the channel code is expected to reduce bit errors but this aspect is not related to our study and well-investigated and characterized in the literature.} To obtain the receiver SNR for S1 under MMSE decoding, we only need the first column of $\mathbf{H}$:
\begin{equation}
\gamma_D=\mathbf{H}_{*,1}^H\mathbf{\Sigma}^{-1}_{\mathbf{w}}\mathbf{H}_{*,1}
\end{equation}
By defining $\gamma=\frac{1}{\sigma^2}$ and elaborating on the above leads to the instantaneous SNR being:
\begin{align}
\label{eqn:snr1}
\gamma_D=\sum^{N_\text{R}}_{r=1}\frac{ g^2_{r}|f_r|^2 |a_{r,1}|^2}{ g^2_{r}|f_r|^2|a_{r,3}|^2+1 }\gamma
\end{align}
From~\eqref{eqn:snr1} we note that decoding at D is conditioned on what has been decoded at the relays. This creates the requirement that the relay must indicate in the preamble of each packet the local SIC decoding results. Since the relays manipulate information at the block level, the SIC results are communicated with the transmitted block on the second hop. Thus, the overhead of indicating the above four local decoding results at each relay for each block they forward is negligible (2 bits for every block). The CSI requirement is tackled as we described earlier.

\section{Outage Analysis}
\label{section:outage-analysis}
In this section we calculate the outage probability of the proposed protocol. We only study the performance of S1 since the same result applies to S2. For minimizing the complexity of our derived expressions, we define the following random variables along with their expectation: $Y$=$|h_{1,1}|^2,X$=$|h_{2,1}|^2,\Exp[|h_{1,1}|^2]$=$\frac{1}{\mu_1},\Exp[|h_{2,1}|^2]$=$\frac{1}{\lambda_1}$, and the same for every relay $r$. For the second hop we only need the average channels $\Exp[|f_{r}|^2]$=$\frac{1}{\nu_r}$, as we will see. Thus, we only manipulate exponential random variables in the remaining parts of this section. Also, in this analysis we consider error propagation in SIC which is critical for its proper performance modeling. Considering complex Gaussian codebooks, for a channel code at $R$ bits/symbol/Hz, information is lost when the instantaneous channel capacity is lower than $R$, leading to the outage probability $\text{$\text{Pout}$}$=$\Pr\{I(|h|^{2}SNR)< R\}$ where $I(x)$=$\log_2(1+x)$ is the Gaussian channel capacity and $|h|$ is the fading amplitude which is the same for the entire block.

\textbf{Outage Probability of Joint Events.} With our protocol, the outage probability of the transmissions from S1 has to be calculated for all the decoding events at the $N_\text{R}$ relays. This is necessary since our protocol adapts the transmitted signals on the second hop depending on what it is decoded at the relays. The end-to-end outage probability expression has to consider 4$^{N_\text{R}}$ potential decoding events. However, we notice that because the random complex channel gains between the sources and the relays are independent, the SIC decoding events at the relays are also independent. To capture formally the above, let $I_{ir}$ be the instantaneous mutual information between the signal transmitted from source $i$ and the received at relay $r$. This allows us to write for the probability of the representative event that no block is decoded at all relays:
\begin{align}
\label{eqn:Pout12zero}
& \Pr\{I_{11}<\frac{3}{2}R_1,I_{21}<\frac{3}{2}R_2,I_{12}<\frac{3}{2}R_1,I_{22}<\frac{3}{2}R_2,...\}\nonumber\\
&=\prod^{N_\text{R}}_{r=1} \Pr\{I_{1r}<\frac{3}{2}R_1,I_{2r}<\frac{3}{2}R_2\}
\end{align}

One can similarly write the probability of all the other 4$^{N_\text{R}}$-1 events. The end-to-end outage probability expression that is provided next is partially presented due to its several terms:
\begin{align}
\label{eqn:outage_e2e}
&\text{Pout}_{\text{S1}}=\prod^{N_\text{R}}_{r=1} \Pr\{I_{1r}<\frac{N_\text{F}}{2}R_1,I_{2r}<\frac{N_\text{F}}{2}R_2\}\nonumber \\
& \times \Pr\{\log_2 (1+\gamma_D(\bm{a})<\frac{N_\text{F}}{2}R_1)\}\nonumber \\
&+ \Pr\{I_{11}>\frac{3}{2}R_1,I_{21}<\frac{N_\text{F}}{2}R_2\}\nonumber  \\
&\times  \prod^{N_\text{R}}_{r=2} \Pr\{I_{1r}<\frac{N_\text{F}}{2}R_1,I_{2r}<\frac{N_\text{F}}{2}R_2\}\nonumber\\
& \times \Pr\{\log_2 (1+\gamma_D(\bm{a})<\frac{N_\text{F}}{2}R_1)\}+...
\end{align}
This expression takes into account the probability of a specific decoding result at all the relays, and this is multiplied with the probability that S1 is in outage during the transmission in the second hop. This multiplication is possible since the events on the first and second hops are independent. Since we have two sources and $N_\text{R}$ relays, there are 4$^{N_\text{R}}$ potential decoding outcomes for the two transmitted blocks. Note that the fraction $\frac{N_\text{F}}{2}$ in front of every $R_1,R_2$ is because our protocol transmits 2 blocks over $N_\text{F}=1+N_\text{R}$ time slots.

Another consequence of the result in \eqref{eqn:Pout12zero} is that the outage expressions for the first relay, can be directly re-used for every relay. The only difference in this case will be the different average channel gains (i.e., $\lambda_r,\mu_r$). The conclusion from this discussion is that we only need to calculate the probability of four events at a single relay.

\begin{figure}[t]
\centering
\subfigure[ $R_1=R_2=1$ bit/symbol]{ \includegraphics[keepaspectratio,width = 0.495\linewidth]{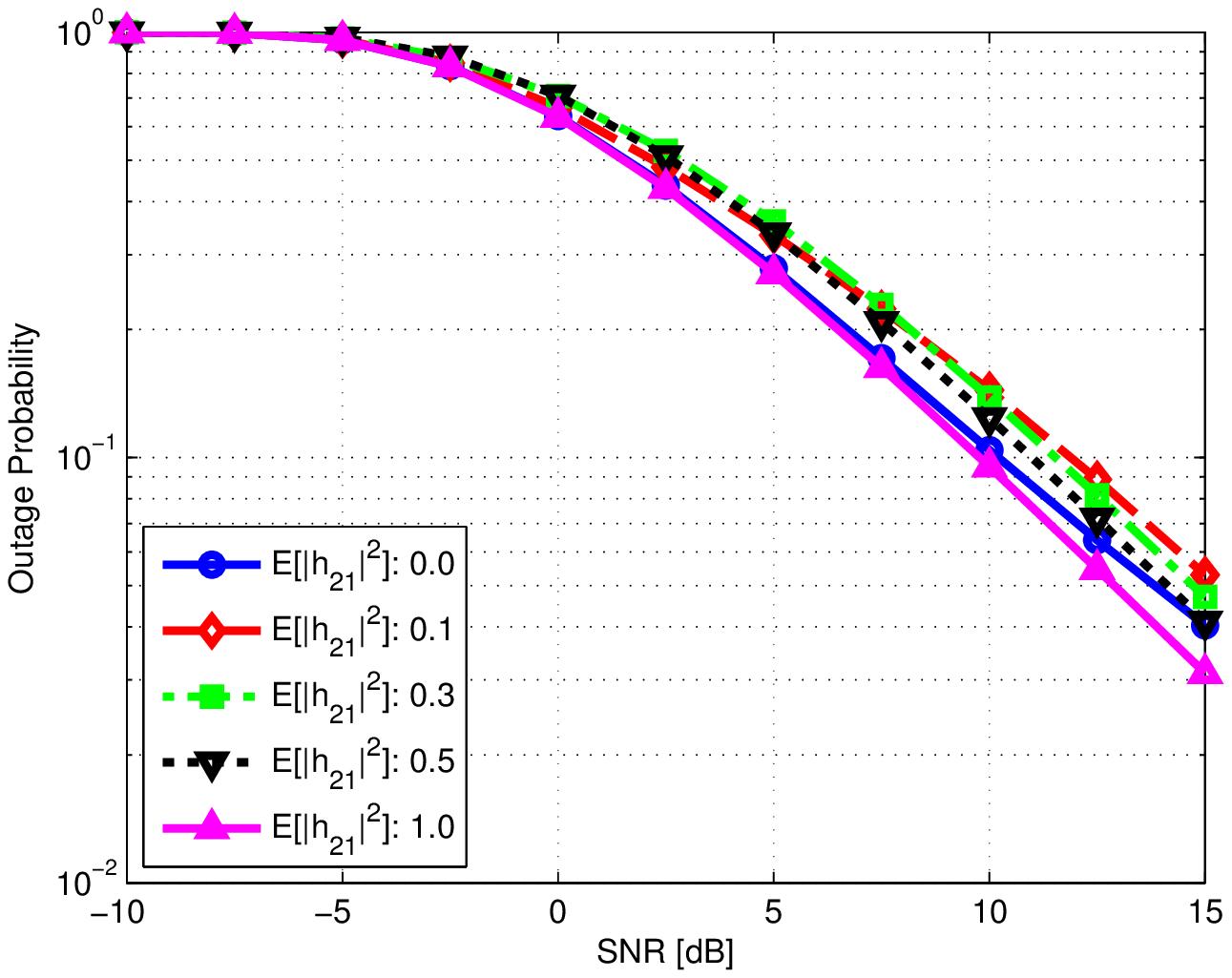}}\hspace{-0.2cm}%
\subfigure[ $R_1=1,R_2=2$ bits/symbol]{ \includegraphics[keepaspectratio,width = 0.495\linewidth]{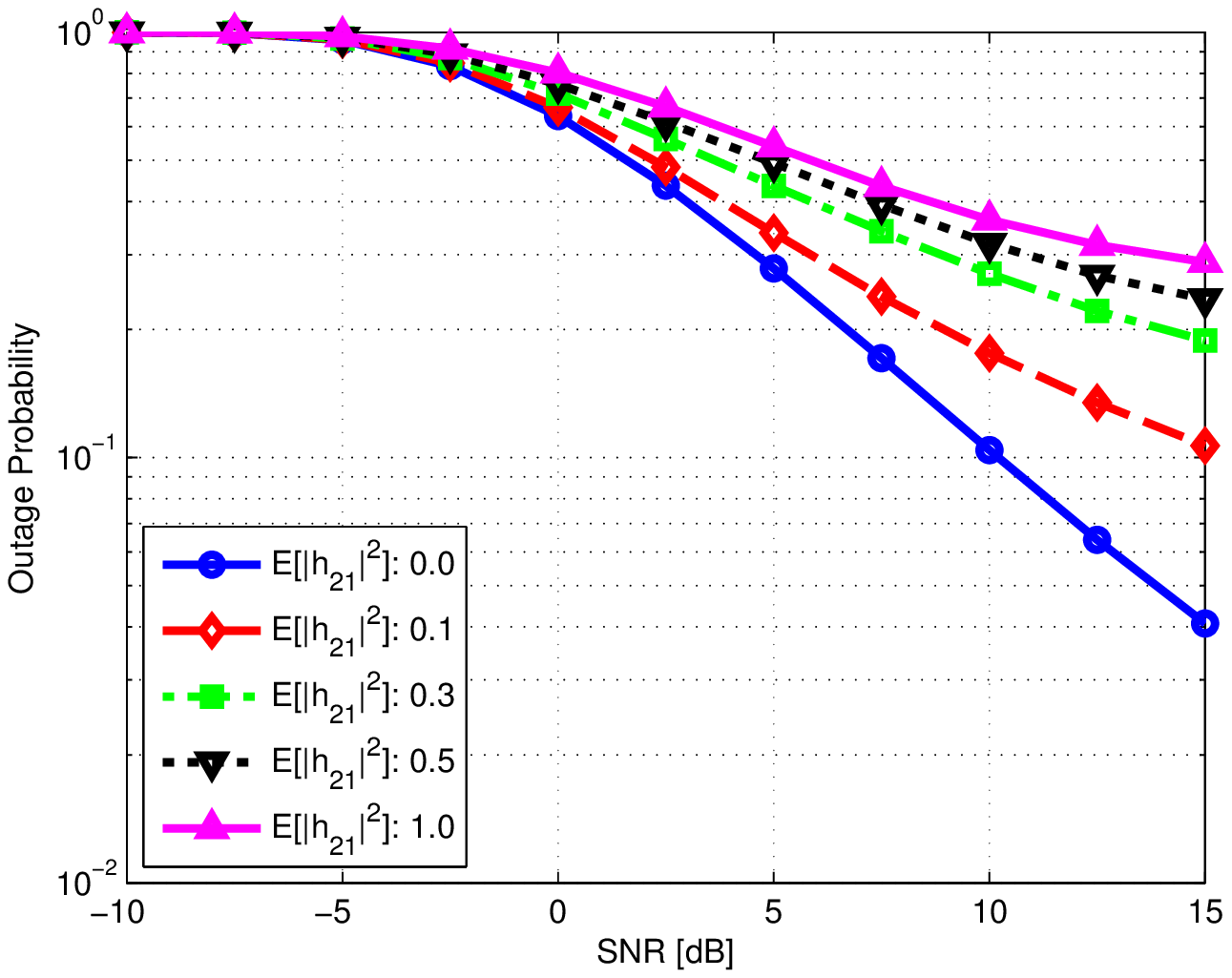}}%
\caption{Numerical outage results for $\Pr\{I_{11}<\frac{3}{2}R_1,I_{21}<\frac{3}{2}R_2)$.}
\label{fig:results2ab}
\end{figure}

The first event is that both users are simultaneously in outage at the first relay:
\begin{equation}
\label{eqn:Pout12a}
\text{Pout}_{\text{S1S2@RS1}}= \Pr \{I_{11}<\frac{N_\text{F}}{2}R_1,I_{21}<\frac{N_\text{F}}{2}R_2\}
\end{equation}
Now the important detail is that if we apply SIC at the relay, $I_{11}$ will be different depending which signal has the highest energy/bit since this will be the one that will be decoded first. In particular if the signal from S1 has higher energy/bit than the signal from S2, i.e., if $\{Y>\frac{k_1}{k_2}X\}$ (this is condition~\eqref{eqn:sic-condition}), it will be $I_{11}$=$\log_2 (1+\frac{Y}{X+\sigma^2})$=$\log_2 (1+\frac{\gamma Y}{\gamma X+1})$, where $\gamma$=$1/\sigma^2$ is the transmit SNR. Here, we have defined $k_1$=$2^{\frac{N_\text{F}}{2}R_1}-1$, $k_2$=$2^{\frac{N_\text{F}}{2}R_2}-1$ as the SNR block decoding thresholds, again in our effort to minimize notation. Now in the opposite case that the signal from S2 has highest energy/bit than S1, then SIC will decode first source S2 which means that it can remove it from the aggregate signal. Thus,
$I_{11}$=$\log_2 (1+\gamma Y)$.
Hence, the event in~\eqref{eqn:Pout12a} can be decomposed to two mutually exclusive events depending on which signal is decoded first:
\begin{align}\label{eqn:Pout12b}
&\Pr\{I_{11}<\frac{N_\text{F}}{2}R_1,I_{21}<\frac{N_\text{F}}{2}R_2\}\nonumber \\
&=\Pr\{I_{11}<\frac{N_\text{F}}{2}R_1,I_{21}<\frac{N_\text{F}}{2}R_2 , Y>\frac{k_1}{k_2}X\} \nonumber \\
&+\Pr\{I_{11}<\frac{N_\text{F}}{2}R_1,I_{21}<\frac{N_\text{F}}{2}R_2 , Y<\frac{k_1}{k_2}X\}
\end{align}
\normalsize
From our previous discussion~\eqref{eqn:Pout12b} can be written as:
\small
\begin{align}\label{eqn:Pout12c}
& \Pr\{I_{11}<\frac{N_\text{F}}{2}R_1,I_{21}<\frac{N_\text{F}}{2}R_2\}= \Pr\{ Y-k_1X<\frac{k_1}{\gamma} ,Y>\frac{k_1}{k_2}X\} \nonumber\\
&+ \Pr\{ X-k_2Y < \frac{k_2}{\gamma} ,Y<\frac{k_1}{k_2}X\}
\end{align}
\normalsize
The event $\{Y>\frac{k_1}{k_2}X\}$ considers all the cases where the block $x_1$, that originates from S1, has the highest energy/bit and thus it will be decoded first. If this last event is true, then $\{Y-k_1X<\frac{k_1}{\gamma}\}$ is the event that S1 cannot be decoded (event $\{I_{11}<\frac{N_\text{F}}{2}R_1\}$). Note that this joint event includes the case that S2 cannot be decoded if $\{Y>\frac{k_1}{k_2}X\} $ is true. The reason is simply that the energy/bit is lower for S2 and so if S1 cannot be decoded we cannot decode source S2. On the other hand, when $\{Y<\frac{k_1}{k_2}X\}$ is true in the second of the two independent events in~\eqref{eqn:Pout12c}, then similarly with before we only need to consider the probability that S2 will be in outage since in this case S1 will definitely be in outage.

To calculate~\eqref{eqn:Pout12c} recall that $X$ and $Y$ are independent exponential random variables. This means that their joint probability density function (PDF) is separable. Thus, for the first event in~\eqref{eqn:Pout12c}:
\begin{align}\label{eqn:Pout12d}
& \Pr \{ Y-k_1X<\frac{k_1}{\gamma} ,Y>\frac{k_1}{k_2}X\} = \int\limits_{0}^{\infty}\int\limits_{x\frac{k_1}{k_2}}^{k_1x+\frac{k_1}{\gamma}}f_X(x)f_Y(y)dydx \nonumber\\
&=\frac{\lambda_1}{\lambda_1+\mu_1 \frac{k_1}{k_2}}- \frac{\lambda_1 \exp(-\frac{\mu_1 k_1}{ \gamma})}{\lambda_1+\mu_1 k_1}
\end{align}
\normalsize
Similarly we calculate the probability of the second event in~\eqref{eqn:Pout12c} and then by adding the two results we finally obtain:
\small
\begin{align}\label{eqn:Pout12e}
& \text{Pout}_{\text{S1S2@RS1}}=\Pr\{I_{11}<\frac{N_\text{F}}{2}R_1,I_{21}<\frac{N_\text{F}}{2}R_2\}\nonumber\\
&= \frac{\lambda_1}{\lambda_1+\mu_1 \frac{k_1}{k_2}}- \frac{\lambda_1 \exp(-\frac{\mu_1 k_1}{\gamma})}{\lambda_1+\mu_1 k_1} + \frac{\mu_1}{\lambda_1  \frac{k_2}{k_1}+\mu_1}- \frac{\mu_1 \exp(-\frac{\lambda_1 k_2}{\gamma})}{\mu_1+\lambda_1 k_2}
\end{align}
\normalsize
Before we proceed, it is instructive to understand an extreme case where $\lambda_1 \rightarrow \infty$. This corresponds to a very small value for the average channel gain from S2 to RS1, i.e., $\Exp [|h_{2,1}|^2]\rightarrow 0$. This leads to $\text{Pout}_{\text{S1S2@RS1}}\rightarrow 1-\exp (-\frac{\mu_1 k_1}{\gamma })$ as expected. By setting the other extreme value $\lambda_1 \rightarrow 0$ or $\Exp [|h_{2,1}|^2]\rightarrow \infty$ the outage probability becomes zero. Now for different rate requirements we can see from~\eqref{eqn:P_out_sic_high_SNR} that the outage event does not converge to zero in the high SNR regime.

With a similar methodology we calculate the probability of the other three events at the first relay. In particular the second event considers the case that S1 in outage while S2 is not:
\begin{align*}
 &\text{Pout}_{\text{S1@RS1}} =\Pr \{I_{11}<\frac{N_\text{F}}{2}R_1,I_{21}>\frac{N_\text{F}}{2}R_2\} \nonumber \\
 &= \Pr \{I_{11}<\frac{N_\text{F}}{2}R_1,I_{21}>\frac{N_\text{F}}{2}R_2, Y>\frac{k_1}{k_2}X\}\\
&+\Pr \{I_{11}<\frac{N_\text{F}}{2}R_1,I_{21}>\frac{N_\text{F}}{2}R_2, Y<\frac{k_1}{k_2}X\}
\end{align*}
\normalsize
In the last decomposed expression we followed the same procedure with before. Only in this case the probability of the first of the two disjoint events is zero. This is again a result of the behavior of SIC that selects to decode first the symbol with the highest energy/bit: If the signal from S1 is the strongest (i.e., \{$Y>\frac{k_1}{k_2}X\}$), and the destination fails to decode it (i.e., $\{I_{11}<\frac{N_\text{F}}{2}R_1\}$), then we cannot decode S2 since the signal from S1 was not cancelled. This leads to:
\begin{align}\label{eqn:Pout1b}
 \text{Pout}_{\text{S1@RS1}} & = \Pr \{I_{11}<\frac{N_\text{F}}{2}R_1,I_{21}>\frac{N_\text{F}}{2}R_2, Y<\frac{k_1}{k_2}X\}\nonumber \\
                           & = \Pr \{Y< \frac{k_1}{\gamma},X-k_2Y > \frac{k_2}{\gamma},Y<\frac{k_1}{k_2}X\}\nonumber \\
                           & = \frac{\mu_1 \exp(- \frac{\lambda_1 k_2}{\gamma})}{\mu_1+\lambda_1 k_2} \Big ( 1-\exp(-\frac{(\mu_1+\lambda_1 k_2)k_1 }{\gamma})\Big )
\end{align}
\normalsize
In the above result note that when the decoding of S2 succeeds, then the signal $x_2$ will be removed from the aggregate. This means that the event that S1 is not decoded is $\{Y<\frac{k_1}{\gamma} \}$ since the decoder must only combat the noise after cancellation. Consider again the case $\lambda_1 \rightarrow \infty$. This corresponds to a very small value for the average channel gain, i.e., $\Exp [|h_{2,1}|^2]\rightarrow 0$  and so the power level of the received signal from S2 is very low. But one notes that we now consider the event $\{Y<\frac{k_1}{k_2}X\}$, i.e., the event that the signal from S1 will be received at an even lower power level than S2. Thus, the probability of the event in \eqref{eqn:Pout1b} becomes zero simply because this event occurs infrequently as $\lambda_1 \rightarrow \infty$.

The third event is symmetric to what we just analyzed: The second source S2 is in outage while the first source S1 is successfully decoded. This is expressed as:
\begin{align}
\text{Pout}_{\text{S2@RS1}} &= \Pr \{ I_{11}>\frac{N_\text{F}}{2}R_1,I_{21}<\frac{N_\text{F}}{2}R_2\}\\
&=\Pr \{ Y-k_1X>\frac{k_1}{\gamma},X < \frac{k_2}{\gamma},Y>\frac{k_1}{k_2}X\} \nonumber\\
&+ \Pr \{I_{11}>\frac{N_\text{F}}{2}R_1,I_{21}<\frac{N_\text{F}}{2}R_2,Y<\frac{k_1}{k_2}X\}\nonumber
\end{align}
\normalsize
With the reasoning we followed in the previous paragraphs we can easily see that the probability of the second of the two events above is zero. Thus, the previous becomes:
\begin{align}\label{eqn:Pout2b}
&\Pr  \{ I_{11}>\frac{N_\text{F}}{2}R_1,I_{21}<\frac{N_\text{F}}{2}R_2\} \nonumber \\
&=\Pr  \{ Y-k_1X>\frac{k_1}{\gamma},X < \frac{k_2}{\gamma},Y>\frac{k_1}{k_2}X\}
\end{align}
\normalsize
The integration is slightly more complicated in this case, and so we calculate~\eqref{eqn:Pout2b} as:
\begin{align}
&\Pr  \{ Y-k_1X>\frac{k_1}{\gamma},X <\frac{k_2}{\gamma},Y> \frac{k_1}{k_2}X\}\nonumber \\
&=\Pr \{ X < \frac{k_2}{\gamma},Y>\frac{k_1}{k_2}X\}\nonumber\\
&-\Pr \{Y-k_1X<\frac{k_1}{\gamma},X <\frac{k_2}{\gamma},Y>\frac{k_1}{k_2}X\}
\end{align}
\normalsize
The calculation gives:
\small
\begin{align}\label{eqn:Pout2c}
&\text{Pout}_{\text{S2@RS1}} =\Pr \{ Y-k_1X>\frac{k_1}{\gamma},X < \frac{k_2}{\gamma},Y> \frac{k_1}{k_2}X\} \nonumber \\ &=\int\limits_{0}^{\frac{k_2}{\gamma}}\int\limits_{x\frac{k_1}{k_2}}^{\infty}f_X(x)f_Y(y)dydx  
-\int\limits_{0}^{\frac{k_2}{\gamma}}\int\limits_{x\frac{k_1}{k_2}}^{k_1x+\frac{k_1}{\gamma}}f_X(x)f_Y(y)dydx \nonumber\\
 &= \frac{\lambda_1 \exp(-\frac{\mu_1 k_1 }{\gamma})}{\lambda_1+\mu_1 k_1} \Big (1-\exp(-\frac{(\lambda_1+\mu_1 k_1)k_2}{\gamma})\Big )  
\end{align}
\normalsize
Even though this event is symmetric to what we analyzed in the previous paragraph, the results and their interpretation is not. To understand the performance
consider again the first case where $\lambda_1 \rightarrow \infty$. This corresponds to a very low value for the average channel gain, i.e., $\Exp [|h_{2,1}|^2]\rightarrow 0$. This leads to $\text{Pout}_{\text{S2@RS1}}\rightarrow \exp (-\frac{\mu_1 k_1}{\gamma})$. Note that this is essentially the case that the received signal from S2 has negligible power and it practically does not interfere. Hence, this means that the outage behavior of S1 approaches the well-known behavior of the point-to-point channel (recall that the outage probability of the event $\{I_{11}<\frac{N_\text{F}}{2}R_1\}$ in this case is $1-\exp (-\frac{\mu_1 k_1}{\gamma})$). Therefore, the probability of the opposite event, i.e., $\{I_{11}>\frac{N_\text{F}}{2}R_1\}$ is $\exp (-\frac{\mu_1 k_1 }{\gamma})$. Care must be given to the high SNR approximation in this case. The high SNR approximation for~\eqref{eqn:Pout2c} with $\lambda_1 \rightarrow \infty$ leads to $\lim_{\gamma\rightarrow \infty} \text{Pout}_{\text{S2@RS1}}= 1$. This is easy to understand because this case corresponds to S2 not being decoded at RS1 which will be true when $\lambda_1 \rightarrow \infty$ (low inter-user interference). However, the high SNR approximation for~\eqref{eqn:Pout2c} with bounded $\lambda_1$ and $\gamma\rightarrow \infty$ leads to $\lim_{\gamma\rightarrow \infty} \text{Pout}_{\text{S2@RS1}}= 0$.

\begin{figure}[t]
\centering
\subfigure[$\Pr (I_{11}<R_1,I_{21}>R_2)$]{ \includegraphics[keepaspectratio,width = 0.49\linewidth]{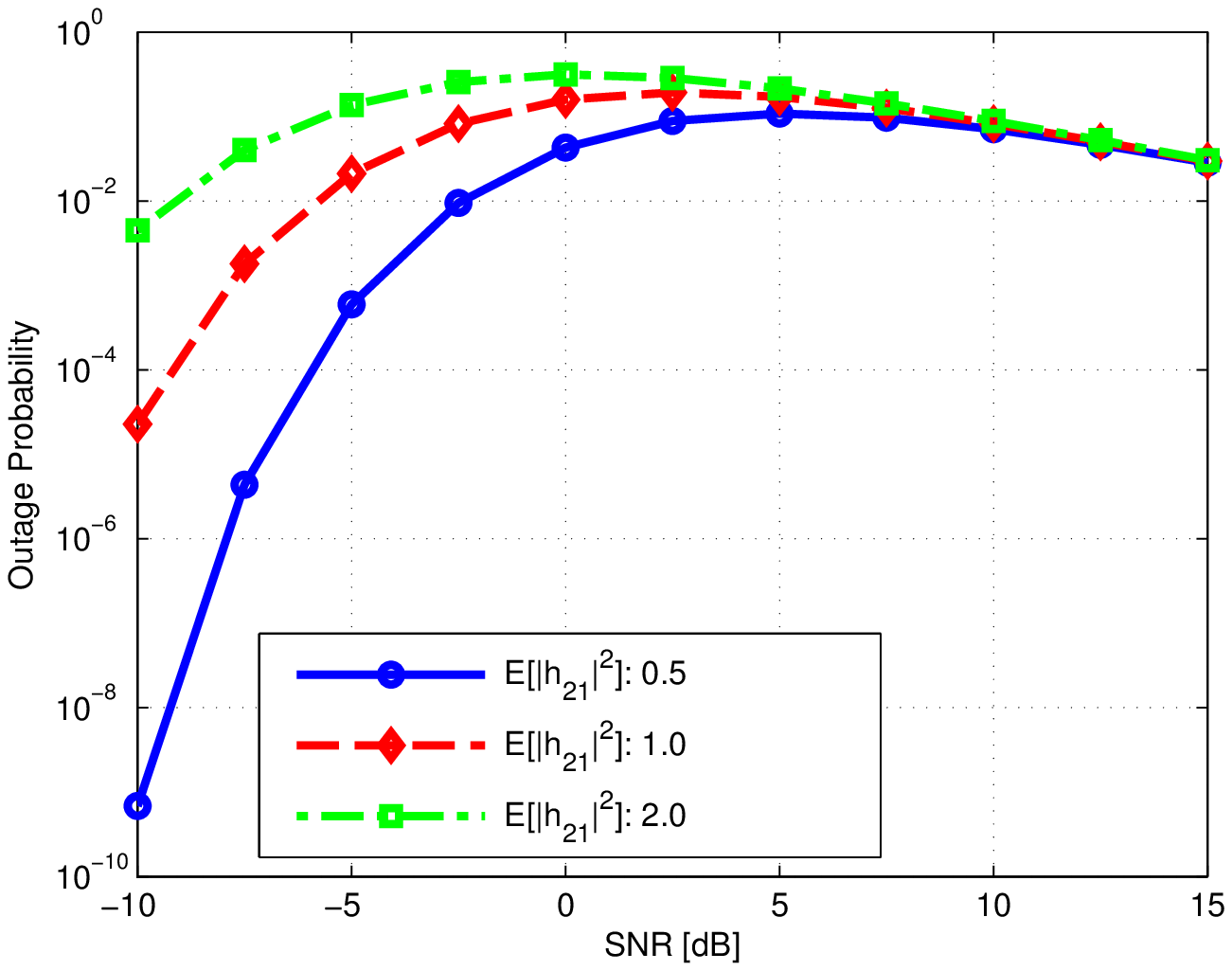}}\hspace{-0.35cm}
\subfigure[$\Pr (I_{11}>R_1,I_{21}<R_2)$]{ \includegraphics[keepaspectratio,width = 0.49\linewidth]{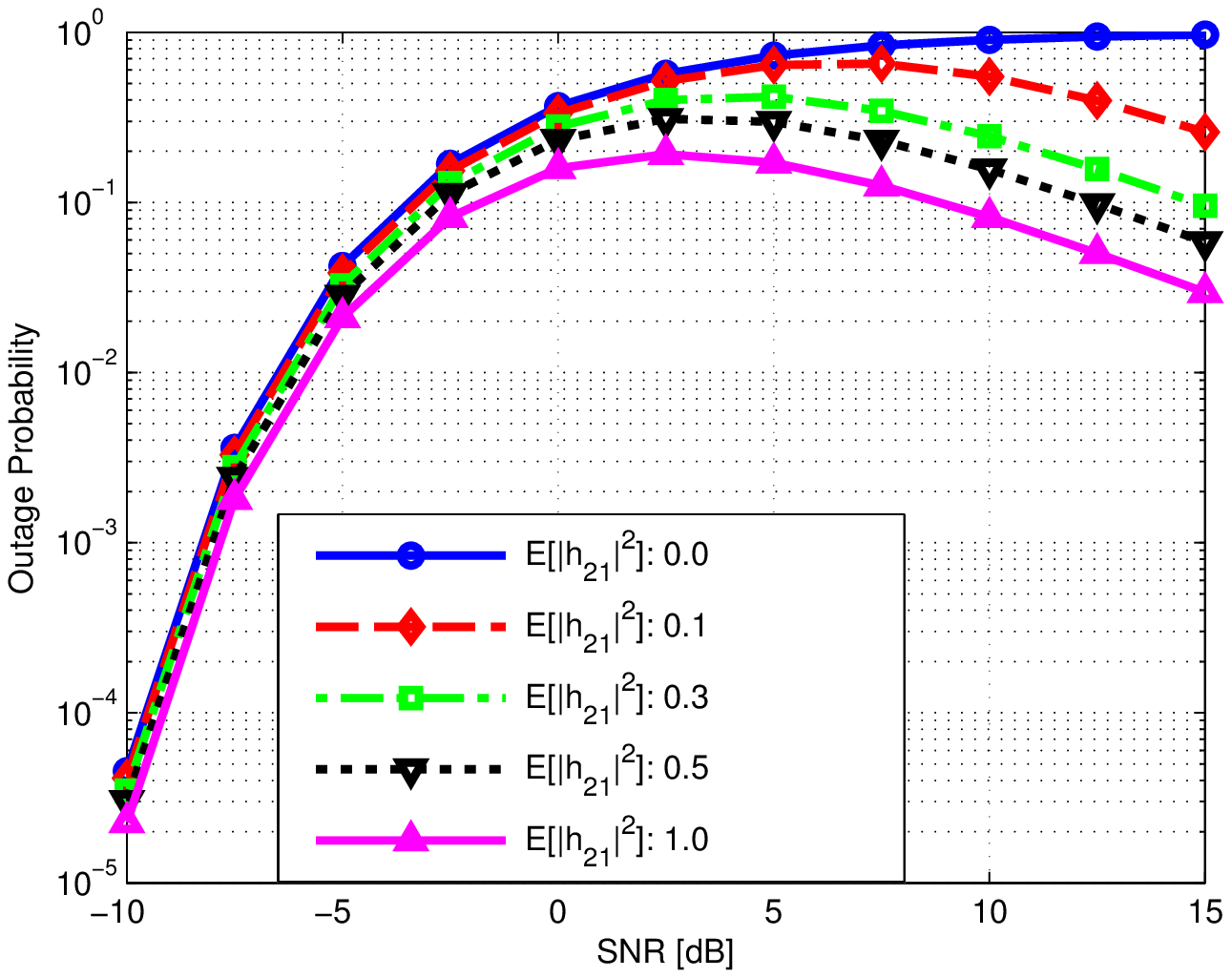}}
\caption{Numerical outage results of the joint events.}
\label{fig:results5}
\end{figure}

Finally, we consider the event that both sources are not in outage at the first relay. Since all the four events that we described in the last few paragraphs are mutually exclusive, and these probability calculations sum up to one, we can easily derive this last quantity. It is thus:

\small
\begin{align*}
\text{Pout}_{\text{none@RS1}} &=\Pr \{ I_{11}>\frac{N_\text{F}}{2}R_1,I_{21}>\frac{N_\text{F}}{2}R_2\}\nonumber\\
&=\Pr \{ Y-k_1X>\frac{k_1}{\gamma},X > \frac{k_2 }{\gamma},Y>\frac{k_1}{k_2}X\})\nonumber\\
&+\Pr \{ Y>\frac{k_1 }{\gamma},X-k_2Y > \frac{k_2 }{\gamma},Y<\frac{k_1}{k_2}X\}\nonumber
\end{align*}
\normalsize
And this becomes:
\begin{align}\label{eqn:Poutn}
\text{Pout}_{\text{none@RS1}} &= 1+\frac{\lambda_1 \exp(-\frac{\mu_1 k_1}{\gamma})}{\lambda_1+\mu_1 k_1} \exp(-\frac{(\lambda_1+\mu_1 k_1)k_2}{\gamma})\nonumber\\
&+ \frac{\mu_1 \exp(-\frac{\lambda_1 k_2}{\gamma})}{\mu_1+\lambda_1 k_2} \exp(-\frac{(\mu_1+\lambda_1 k_2)k_1}{\gamma})\nonumber\\
&-\frac{\lambda_1}{\lambda_1+\mu_1 \frac{k_1}{k_2}}-\frac{\mu_1}{\mu_1+\lambda_1 \frac{k_2}{k_1}}
\end{align}
As we explained initially in this section, the outage expressions for the $r$-th relay are exactly the same. The only difference in this case are the average channel gains.

\textbf{End-to-End Outage Probability.} Now we get back to the main outage expression in~\eqref{eqn:outage_e2e} that includes the outage probability expressions for the second hop. Depending on SIC, different signals are transmitted. For each one of the 4$^{N_\text{R}}$ decoding events, we set all the different values for $a$ in the parameterized SNR expression of $\gamma_D$ given in~\eqref{eqn:snr1}, and then we include the results in the final outage expression in~\eqref{eqn:outage_e2e}. It is challenging to derive closed-form expressions for the second hop, except of course the trivial case where both blocks are decoded at every relay. The reason is that $\gamma_D$ in~\eqref{eqn:snr1} is a fractional expression with exponential random variables in the numerator and the denominator. Nevertheless, we can calculate numerically the outage probability for the second hop and this is what we do when we present the numerical results similar to the related work~\cite{yao12}.

\section{DMT Analysis}
\label{section:dmt}
Two-hop single-user cooperative protocols that transmit a single block can achieve a multiplexing gain of $\frac{1}{2}$~\cite{laneman01}. Our communication scheme uses $N_\text{F}$=$N_\text{R}+1$ slots for transmitting two blocks. So this leads potentially to maximum multiplexing gain of $\frac{2}{3}$ for $N_\text{R}$=2. To get the DMT expression, and verify the previous result, we have to consider all the 4$^{N_\text{R}}$ product terms in~\eqref{eqn:outage_e2e}. In this case we observe that every term in the summation of $\text{Pout}_{\text{S1}}$ will be the product of $N_\text{R}+1$ probability terms (for example $\prod^{N_\text{R}}_{r=1}$Pout$_\text{S1S2@RSr}$Pout$_\text{S1@D}$).

\textbf{First hop.} We calculate first the individual high SNR approximations for the outage probabilities of the first hop. The high SNR approximation for~\eqref{eqn:Pout12e} leads to a constant, i.e., $\lim_{\gamma\rightarrow \infty} \text{Pout}_{\text{S1S2@RS1}}$=$C_1$, with

\small
\begin{equation}\label{eqn:P_out_sic_high_SNR}
C_1= \frac{\lambda_1}{\lambda_1+\mu_1 \frac{k_1}{k_2}}- \frac{\lambda_1}{\lambda_1+\mu_1 k_1}+ \frac{\mu_1}{\lambda_1  \frac{k_2}{k_1}+\mu_1}- \frac{\mu_1 }{\mu_1+\lambda_1 k_2}.
\end{equation}
\normalsize
Similarly we obtain for RSr that $ \lim_{\gamma\rightarrow \infty} \text{Pout}_{\text{S1S2@RSr}}$=$C_r$. The intermediate cases luckily lead to zero terms. For example the high SNR approximation for~\eqref{eqn:Pout1b} leads to $\lim_{\gamma\rightarrow \infty} \text{Pout}_{\text{S1@RS1}}$=0. Similarly it can be verified that $\lim_{\gamma\rightarrow \infty} \text{Pout}_{\text{S2@RS1}}$=0 and of course the same expressions are zero for any other relay RSr.
The high SNR approximation of~\eqref{eqn:Poutn} leads to another constant $\lim_{\gamma\rightarrow \infty} \text{Pout}_{\text{none@RS1}}=C^{'}_1$, where
\begin{equation}\label{eqn:P_out_sic}
C^{'}_1= 1+\frac{\lambda_1}{\lambda_1+\mu_1 k_1}+\frac{\mu_1}{\mu_1+\lambda_1 k_2}-\frac{\lambda_1}{\lambda_1+\mu_1 \frac{k_1}{k_2}}
-\frac{\mu_1}{\mu_1+\lambda_1 \frac{k_2}{k_1}}.
\end{equation}
\normalsize
Similarly we obtain for RSr that $ \lim_{\gamma\rightarrow \infty} \text{Pout}_{\text{none@RSr}}$=$C^{'}_r$. The above analysis means that in the high SNR regime we only have to consider $2^{N_\text{R}}$ terms in~\eqref{eqn:outage_e2e}, i.e., the cases that either both or none of the signals are decoded at a relay.

\textbf{Second hop.} The best case is that both signals from S1, S2 are decoded at every relay. From~\eqref{eqn:snr1} we have that for S1:
\begin{align}
\label{eqn:PoutUB}
\gamma_D=\frac{\sum^{N_\text{R}}_r g_r^2 |f_r|^2}{\sigma^2}, \lim_{\gamma\rightarrow \infty} \text{Pout}_{\text{S1@D}}\doteq \Big ( \prod^{N_\text{R}}_{r} \nu_r \Big ) \Big ( \frac{2^{\frac{N_\text{F}}{2}R_1}-1}{\gamma}  \Big )^{N_\text{R}}
\end{align}
\normalsize
From~\eqref{eqn:P_out_sic} and~\eqref{eqn:PoutUB} we have that the approximation for this first event is: 
\small
\begin{align}\label{eqn:log_P_out_log_gamma3}
 \lim_{\gamma\rightarrow \infty} \text{Pout}^\text{event1}_{\text{S1}} &= \prod^{N_\text{R}}_{r=1}\text{Pout}_\text{none@RSr} \text{Pout}_{\text{S1@D}}
  \doteq   \Big ( \prod^{N_\text{R}}_{r} C^{'}_r \nu_r \Big ) \Big ( \frac{2^{\frac{N_\text{F}}{2}R_1}-1}{\gamma}\Big )^{N_\text{R}}
\end{align}
\normalsize

The other events are covered by he worst case that no block is decoded at all relays:
\begin{align}\label{eqn:log_P_out_log_gamma4}
\text{Pout}^\text{event2}_{\text{S1}}&=\prod^{N_\text{R}}_{r=1}\text{Pout}_\text{S1S2@RSr} \text{Pout}_{\text{S1@D}}
\end{align}
\normalsize
The first terms converge to constants $C_r$ while we only need to approximate the final term in \eqref{eqn:log_P_out_log_gamma4}. To do that, and to simplify the final result, we create the equivalent channel model for~\eqref{eqn:signal_model1}: We denote the first column vector of $\mathbf{H}$ as $\mathbf{H}_{*,1}=[h_{1,d}f_1~~h_{2,d}f_2~~...~~h_{r,d}f_r]^T$. If the instantaneous channel matrix is $H$, the mutual information between the transmitted signal $x_1$ and the sufficient statistic $\mathbf{\tilde{y}}$ at the receiver is:
\begin{align}
I(x_1;\mathbf{\tilde{y}}|\mathbf{H}=H)&=I(x_1;\mathbf{H}\mathbf{x}+\mathbf{w}|\mathbf{H}=H)\nonumber\\
&\stackrel{\eqref{eqn:test4}}{=}I(x_1;\mathbf{H}_{*,1}x_1+\mathbf{w}|\mathbf{H}=H)\label{eqn:test4}
\end{align}
In the above \eqref{eqn:test4} is because the sufficient statistic for $x_1$ requires only the first column of $\mathbf{H}$. The mutual information of the 1x$N_\text{R}$ MIMO channel $\mathbf{H}_{*,1}x_1+\mathbf{w}$ is:
\begin{equation}
\log \det ( \mathbf{I}_{N_\text{R}}+ \mathbf{H}_{*,1} \mathbf{H}^{H}_{*,1} \mathbf{\Sigma}^{-1}_{\mathbf{w}} )\label{eqn:test3}
\end{equation}
The mutual information can help calculate the behavior of the lower bound on the outage probability. Namely:

\begin{align}
\text{Pout}^\text{LB}_{\text{S1@D}} &= \Pr \{ \log \det ( \mathbf{I}_{N_\text{R}}+ \mathbf{H}_{*,1} \mathbf{H}^{H}_{*,1} \mathbf{\Sigma}^{-1}_{\mathbf{w}} ) \leq \frac{N_\text{F}}{2}R_1 \}\nonumber\\
& \stackrel{\eqref{eqn:test}}{\doteq} \Pr \{ \log \det ( \mathbf{I}_{N_\text{R}}+ \frac{1}{\sigma^2}\mathbf{H}_{*,1} \mathbf{H}^{H}_{*,1} )\leq \frac{N_\text{F}}{2}R_1 \}\label{eqn:test}\\
& = \Pr \{  \log \Big ( 1+\gamma\sum^{N_\text{F}}_{r=1} g^2_{r}|f_r|^2 |h_{r,d}|^2 \Big ) \leq \frac{N_\text{F}}{2}R_1 \}\nonumber\\
& \stackrel{\eqref{eqn:PoutLB}}{\implies} \lim_{\gamma\rightarrow \infty} \text{Pout}^\text{LB}_{\text{S1@D}}\doteq \Big ( \prod^{N_\text{R}}_{r=1} C_r \nu_r \Big ) \Big ( \frac{2^{\frac{N_\text{F}}{2}R_1}-1}{\gamma} \Big )^{N_\text{R}} \label{eqn:PoutLB}
\end{align}
In the above \eqref{eqn:test} is because the colored noise does not affect the DMT analysis, that is approximated with white noise (\S 9,~\cite{tse04}). Also~\eqref{eqn:PoutLB} is because the resulting random variable that is the product of two exponential random variables is independent across the $N_\text{F}$ paths~\cite{jing06}. For the lower bound event2 leads to
\begin{align*}
\lim_{\gamma\rightarrow \infty} \text{Pout}^\text{LB}_{\text{S1@D}}\doteq \Big ( \frac{2^{\frac{N_\text{F}}{2}R_1}-1}{\gamma} \Big )^{N_\text{R}}.
\end{align*}
Finally, the upper bound $\text{Pout}^\text{UB}_{\text{S1@D}}$ is that of~\eqref{eqn:PoutUB}, since it is impossible to have a better behavior for the protocol from the relays to D. So for the complete protocol:
\begin{align*}
\lim_{\gamma\rightarrow \infty} \text{Pout}_{\text{S1}}  \doteq  \Big ( \frac{2^{\frac{N_\text{F}}{2}R_1}-1}{\gamma}\Big )^{N_\text{R}}
\end{align*}
\textbf{Final Result.} Given that the multiplexing gains and the data rates of S1 and S2 are related by: $R_1$=$r_1 \log(\gamma)$,$R_2$=$r_2 \log(\gamma)$,
we have, the DMT expression is:
\begin{equation*}
d=\lim_{\gamma\rightarrow \infty} \frac{-\log \text{Pout}_{\text{S1}}}{\log \gamma}=N_\text{R}-\frac{N_\text{R}N_\text{F}}{2}r_1,0\leq r_1\leq \frac{2}{N_\text{F}}
\end{equation*}
\normalsize
One observes  that the diversity gain of S1 is only affected by $r_1$. This is because S2's mulitplexing gain $r_2$ can only affect S1 in the decoding of S2 with SIC. But whether the relays decode or not, there is no information loss for S1 since AF is used in the later case. This is another key design feature of our protocol, i.e., the use of AF opportunistically is instrumental to avoid diversity loss. Note also that the protocol in~\cite{yao12} achieves the same DMT but for a half-duplex single-hop network, i.e., the high SNR performance is the same. Of course for finite SNR, the performance is different. Both aspects will be examined with simulations.

\section{Optimization in a Multi-Relay Network for Improved SIC Performance}
\label{section:network}
In this part of our work we explore a scenario where from the number of potential relays that are present in the network $N_\text{R}$, a number $N_\text{RU}$ of them is used simultaneously by the proposed protocol. Contrary to classic relay selection protocols that require all relays to overhear the transmitted signal, we propose to perform \textit{relay pre-selection}, i.e., to select statically which set of relays will be used by the protocol for maximizing the SE of the system. Consequently, the maximum diversity gain is $N_\text{RU}$ but the benefit is that by using our model we can pre-select nodes that can have high performance with SIC. The remaining $N_\text{R}-N_\text{RU}$ nodes can remain idle and do not have participate at all in the communication (e.g., no need for overhearing, power consumption) contrary to the behavior of classic cooperative protocols that we reviewed.
To formally define the problem we address now, we use a single optimization variable $x_{r}$ that indicates whether relay $r$ is used. In vector form we have $\bm{x}=\big(x_{r}\in \{0,1\}:r\in\{1,2,...,N_\text{R}\}\big)$. Our objective is to select $N_\text{RU}$ relays so as to maximize the spectral efficiency:

\small
\begin{align}\label{eqn:opt-problem}
&\max_{\bm{x}} \sum_{r\in \mathcal{R}} x_r \Big ( \text{Pout}_{\text{S2@R-r}}(R_1,R_2,\lambda_r,\mu_r)R_1\nonumber\\
&+\text{Pout}_{\text{S1@Rr}}(R_1,R_2,\lambda_r,\mu_r)R_2\nonumber\\
&+\text{Pout}_{\text{none@Rr}}(R_1,R_2,\lambda_r,\mu_r)(R_1+R_2) \Big ) \nonumber\\
&\text{subject to} \sum_{r\in \mathcal{R}} x_{r}=N_\text{RU},x_{r}\in \{0,1\}
\end{align}
\normalsize
The constraint ensures that only $N_\text{RU}$ relays are used. The specific formulation is easily solved in $O(1)$ since we simply select the first $N_\text{RU}$ relays that have the objective with the highest value. Note that this optimization is executed only on large time scales and by considering the average channel. Note that more sophisticated formulations could also include as additional optimization variables both $R_1,R_2$ and also the number of simultaneously used relays $N_\text{RU}$.

\section{Performance Evaluation}
\label{section:performance-evaluation}
Our performance evaluation consists of Monte Carlo simulations and numerical results, and has four objectives: 1) underpin the design choices of our protocol, 2) investigate the performance benefits over related work, 3) validate our performance model, 4) demonstrate the use of our system in a multi-relay network setting. Since the numerical analysis corresponds to the outage probability that assumes the use of a capacity-achieving code, we use an LDPC code with length of 2048 bits in our simulations. We compare our proposed protocol against DSTC protocols that use orthogonal transmissions~\cite{jing06,laneman03}. With DSTC/DF~\cite{laneman03} the sources transmit orthogonally and each relay forwards the decoded, while all the relays broadcast simultaneously with the same Alamouti STC. The DSTC/AF~\cite{jing06} protocol applies the same Alamouti code without decoding the signals. Regarding multi-user (MU) communication protocols we compare with CFNC~\cite{giannakis08} but in the same topology with our proposed protocol (2-hop MAMRN with two sources, two relays, and one destination) to ensure fairness. The ANC-based protocol reported in~\cite{jnl_2010_twc} is also evaluated for the same topology. Finally, we also present the performance of the DF scheme reported in~\cite{yao12} that is denoted in the figures as DSTC/DF1. The important detail with this scheme is that it uses a DF protocol from two sources located one hop away from D. Hence, the comparison is unfair in favor of DSTC/DF1 and against every other tested system, but it is useful to serve as an interesting bound for the maximum multiplexing gain. With the half-duplex protocol in~\cite{yao12}, the two sources transmit independently over two slots, while they overhear the transmission from each other. This means that during a third slot they can both transmit the block they overheard with an Alamouti code. The maximum multiplexing gain is $\frac{2}{3}$, similar to our protocol.

\begin{figure}[t]
\centering
\subfigure[]{\includegraphics[keepaspectratio,width = 0.495\linewidth]{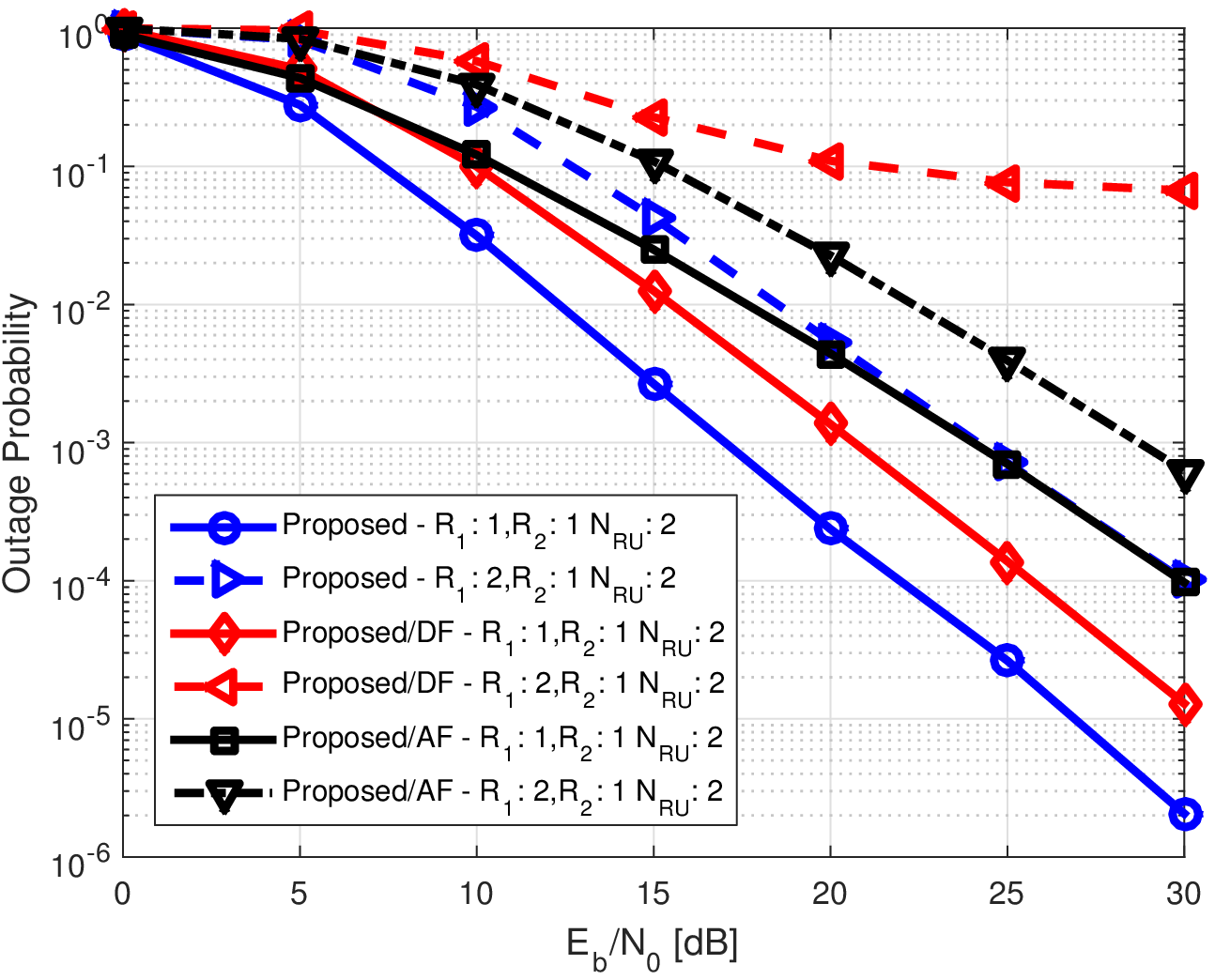}\label{fig:results10}}\hspace{-0.1cm}
\subfigure[]{\includegraphics[keepaspectratio,width = 0.495\linewidth]{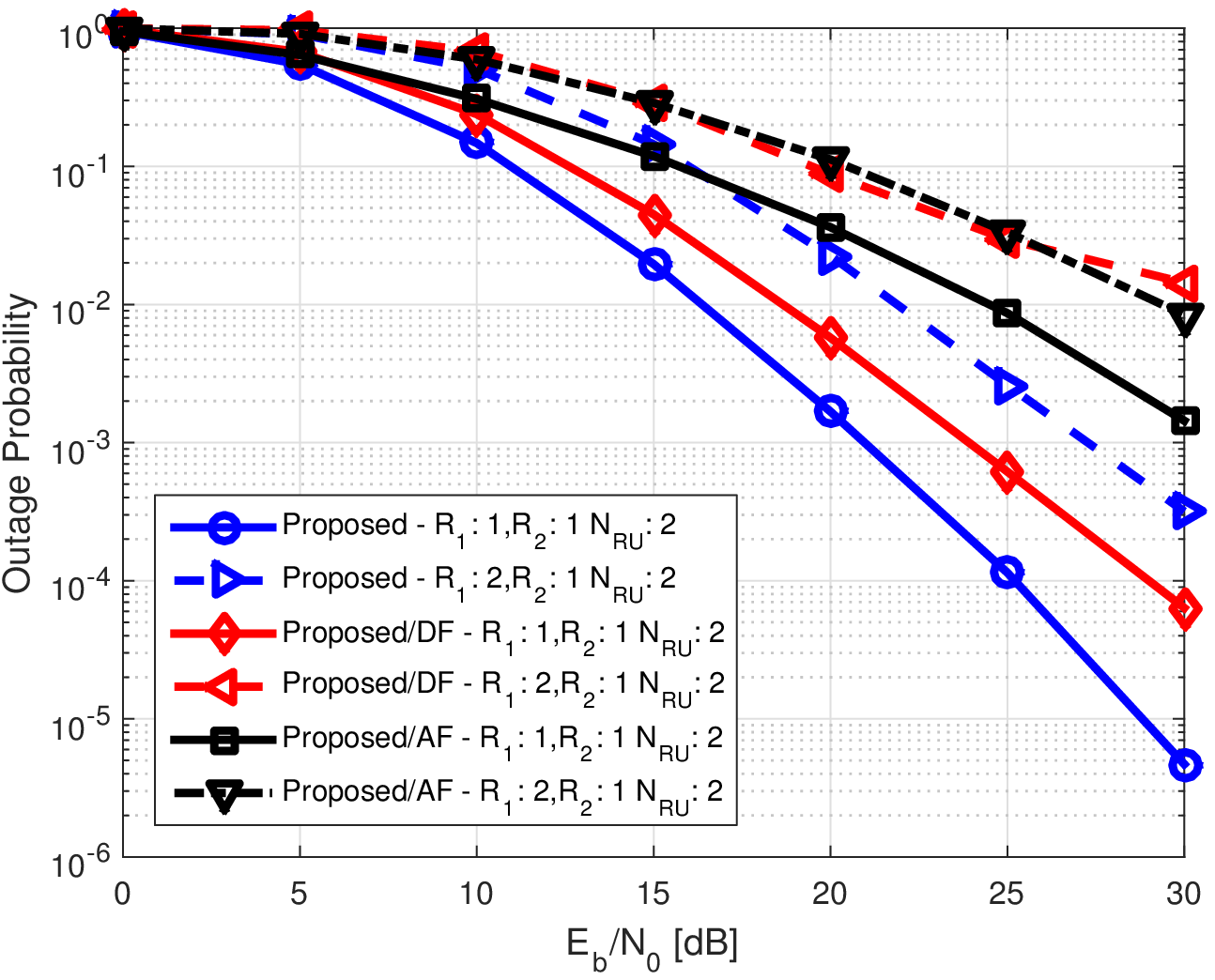}\label{fig:results11}}\hspace{-0.2cm}
\caption{Simulation results for different flavors of the proposed protocol and different $R_1$,$R_2$. a) $\Exp [h^2]$=1 for every channel, and b) $\Exp [h^2_{12}]=\Exp [h^2_{21}]$=0.1 while for the remaining ones $\Exp [h^2]$=1.}
\end{figure}

\begin{figure}[t]
\centering
\subfigure[]{ \includegraphics[keepaspectratio,width = 0.495\linewidth]{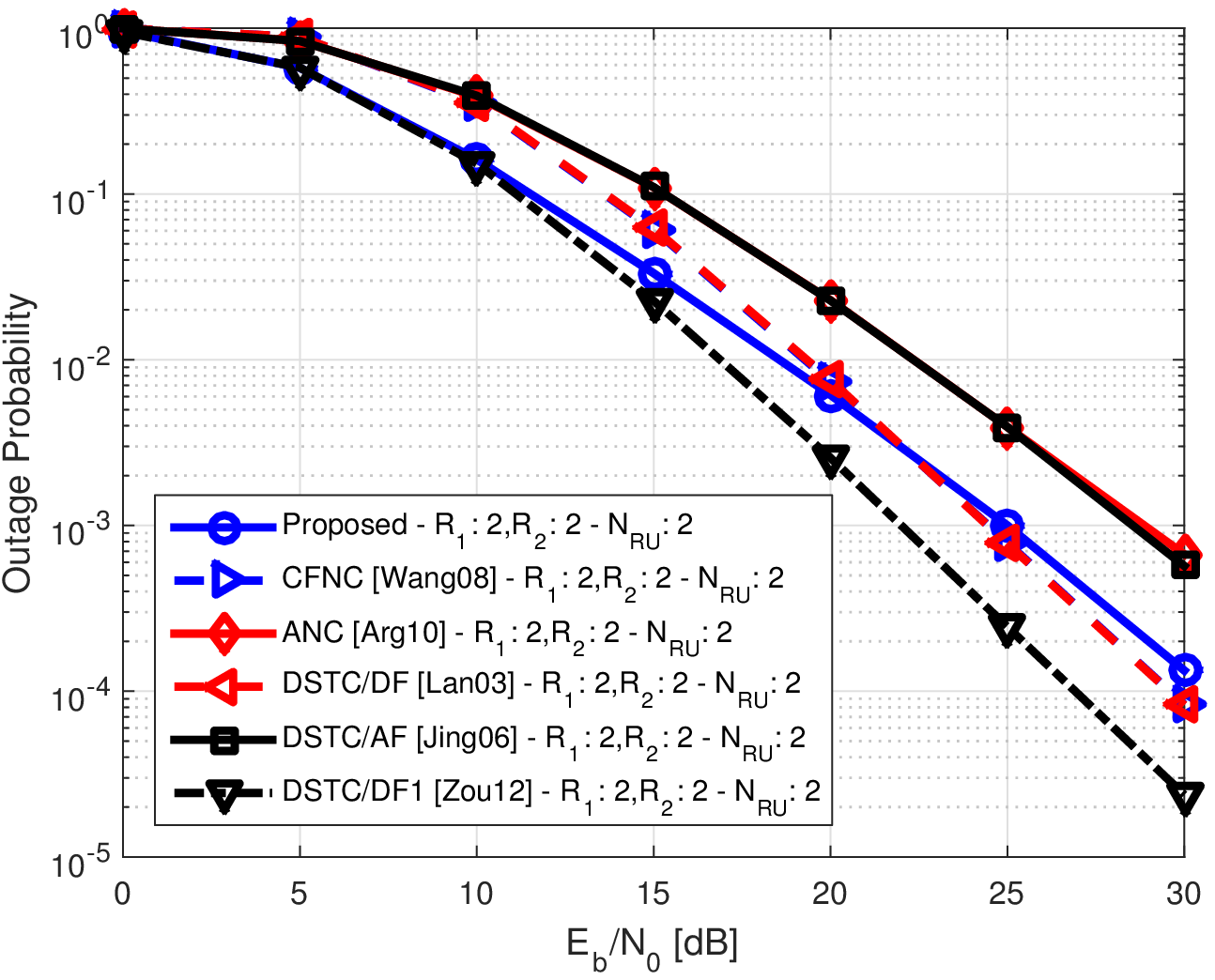}\label{fig:results14}}\hspace{-0.5cm}
\subfigure[]{ \includegraphics[keepaspectratio,width = 0.495\linewidth]{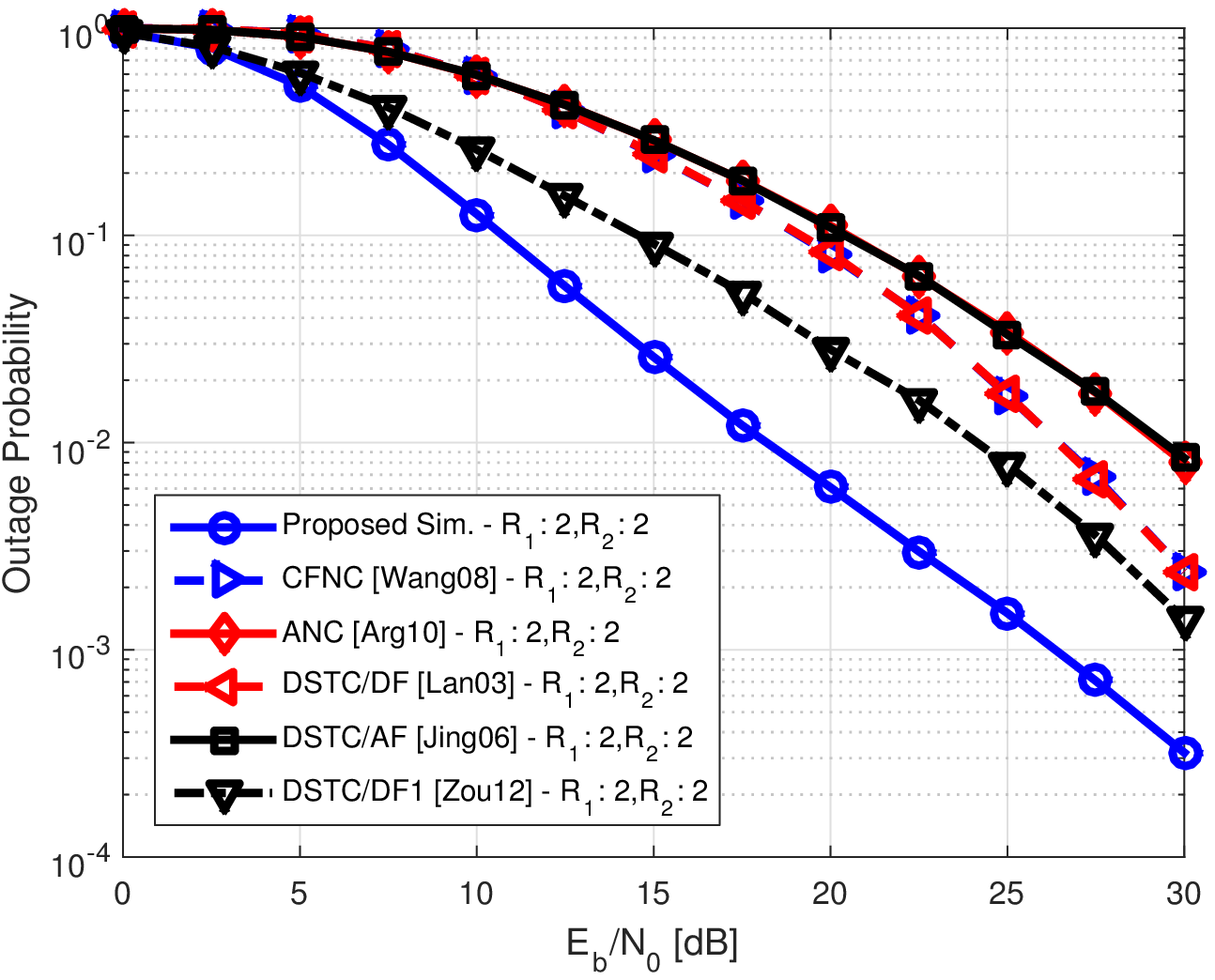}\label{fig:results15}}
\caption{The complete set of the tested protocols for $R_1,R_2$=2 bits/symbol. a) $\Exp [h^2]$=1 for every channel, and b) $\Exp [h^2_{12}]=\Exp [h^2_{21}]$=0.1 while for the remaining ones $\Exp [h^2]$=1. }
\end{figure}

\begin{figure}[t]
\centering
\subfigure[]{ \includegraphics[keepaspectratio,width = 0.495\linewidth]{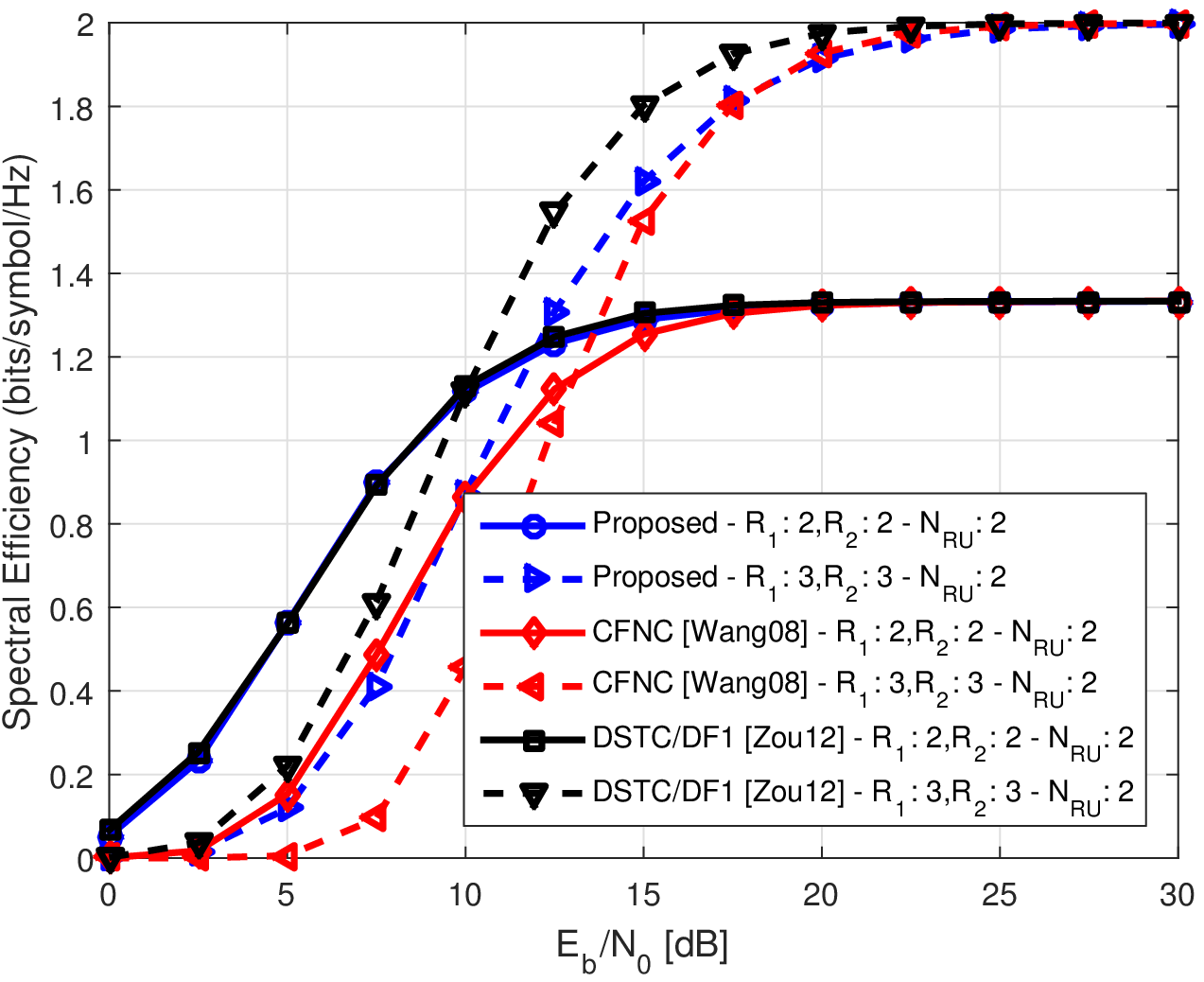}\label{fig:results14_tput}}\hspace{-0.5cm}
\subfigure[]{ \includegraphics[keepaspectratio,width = 0.495\linewidth]{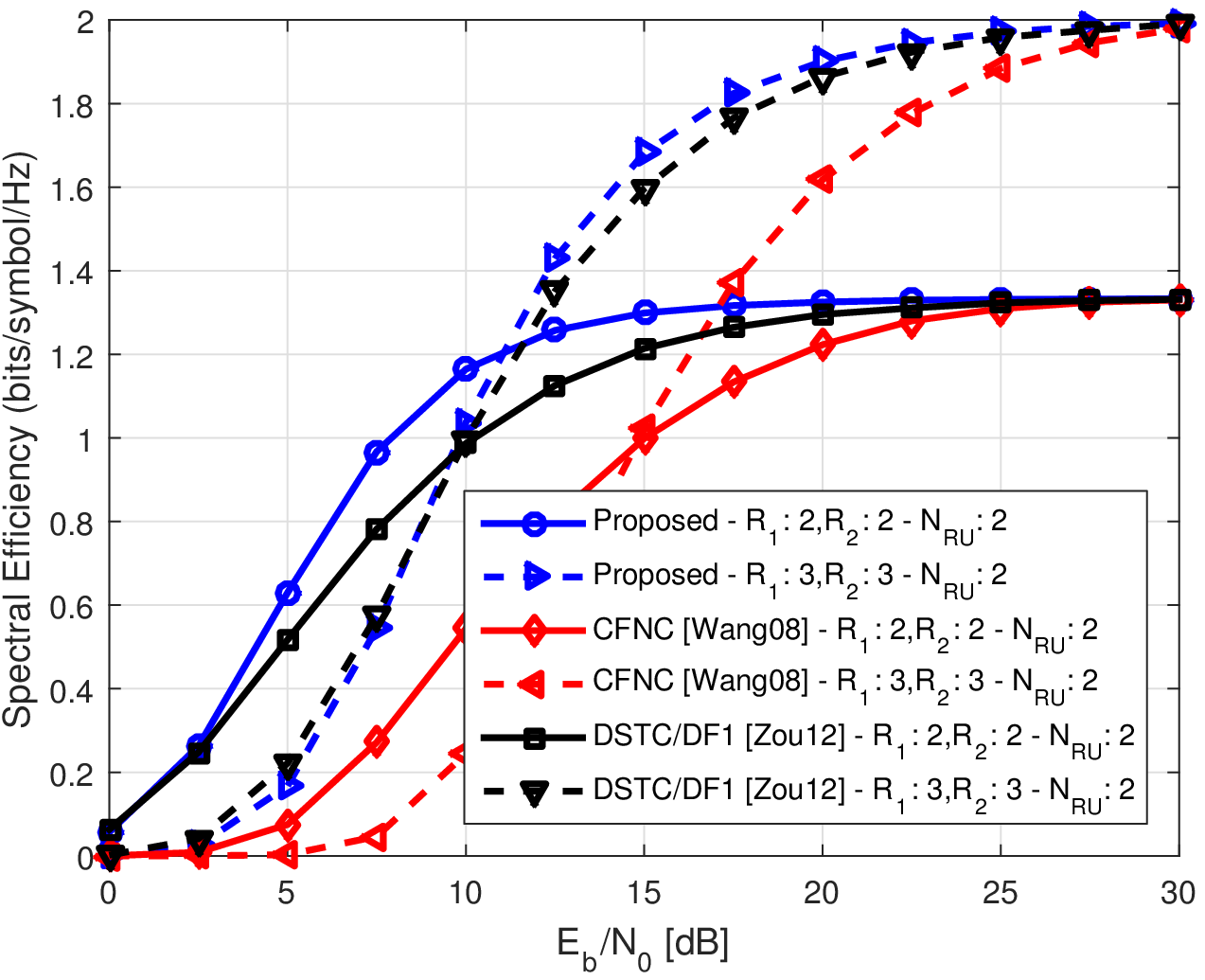}\label{fig:results15_tput}}
\caption{Simulation results for the SE of different protocols and different MCSs. a) $\Exp [h^2]$=1 for every channel, and b) $\Exp [h^2_{12}]=\Exp [h^2_{21}]$=0.1 while for the remaining ones $\Exp [h^2]$=1. }
\end{figure}

\textbf{Protocol Design.}
First we evaluate the fundamental design choices made in our protocol under the presence of two relays. We present two "downgraded" versions of our system namely Proposed/AF that applies forwarding operations without SIC decoding, and Proposed/DF that applies SIC and forwards only the decoded blocks. In Fig.~\ref{fig:results10} we present outage results for a first scenario with high inter-user interference (i.e., $\Exp [|h_{1,2}|^2]$=1, $\Exp [|h_{2,1}|^2]$=1). We note that the Proposed/AF and Proposed/DF are different in terms of performance depending on the MCSs. The reason is that SIC suffers from high inter-user interference and asymmetric MCSs and so the AF strategy is better in this case. For low inter-user interference in Fig.~\ref{fig:results11} the Proposed/DF scheme is slightly better for asymmetric MCSs since it exploits the higher number of decoded blocks, but it is still worse than the Proposed/AF after 25 dB. This makes more clear the need for our protocol that is robust to both low and high inter-user interference and different MCSs, \emph{i.e., it leverages opportunistically decoded signals but when it cannot, the signals are still forwarded avoiding thus information loss.} Hence, a trivial combination of DF or AF strategies with SIC or with undecoded interfering signals respectively, is not enough and robust to different combination of channel gains and MCSs.

\textbf{Outage comparisons for symmetric/asymmetric links.} In Fig.~\ref{fig:results14} we present outage results for a scenario with $N_\text{R}$=$N_\text{RU}$=2 and high inter-user interference (i.e., $\Exp [|h_{1,2}|^2]$=1, $\Exp [|h_{2,1}|^2]$=1). Regarding the protocols that use orthogonality like DSTC/AF~\cite{jing06} and DSTC/DF~\cite{laneman03}, we observe a similar behavior in the sense that DF is better than AF, even though in this case we have orthogonal transmission and there is no inter-user interference. Since the gains on the second relay are strong (i.e., $\Exp [|h_{1,2}|^2]$=1,$\Exp [|h_{2,1}|^2]$=1) the performance is very good for both schemes. For the MU-based protocol CFNC generally performs very well for symmetric links and approaches the performance of DSTC/DF and DSTC/DF1. CFNC extracts the maximum diversity gain since the channel between each source and the two relays is independent and strong. However, our proposed protocol performs better below 20 dB while CFNC is better beyond 20 dB. Our scheme approaches the performance of the orthogonal DSTC/DF~\cite{jing06}. Overall our protocol is inferior in terms of outage only in the high SNR regime only for a fully symmetric network. As we will later see even this is not a problem in terms of SE.

\begin{figure*}[!htb]
\centering
\subfigure[$R_1$=$R_2$=1 bit/symbol]{\includegraphics[keepaspectratio,width = 0.32495\linewidth]{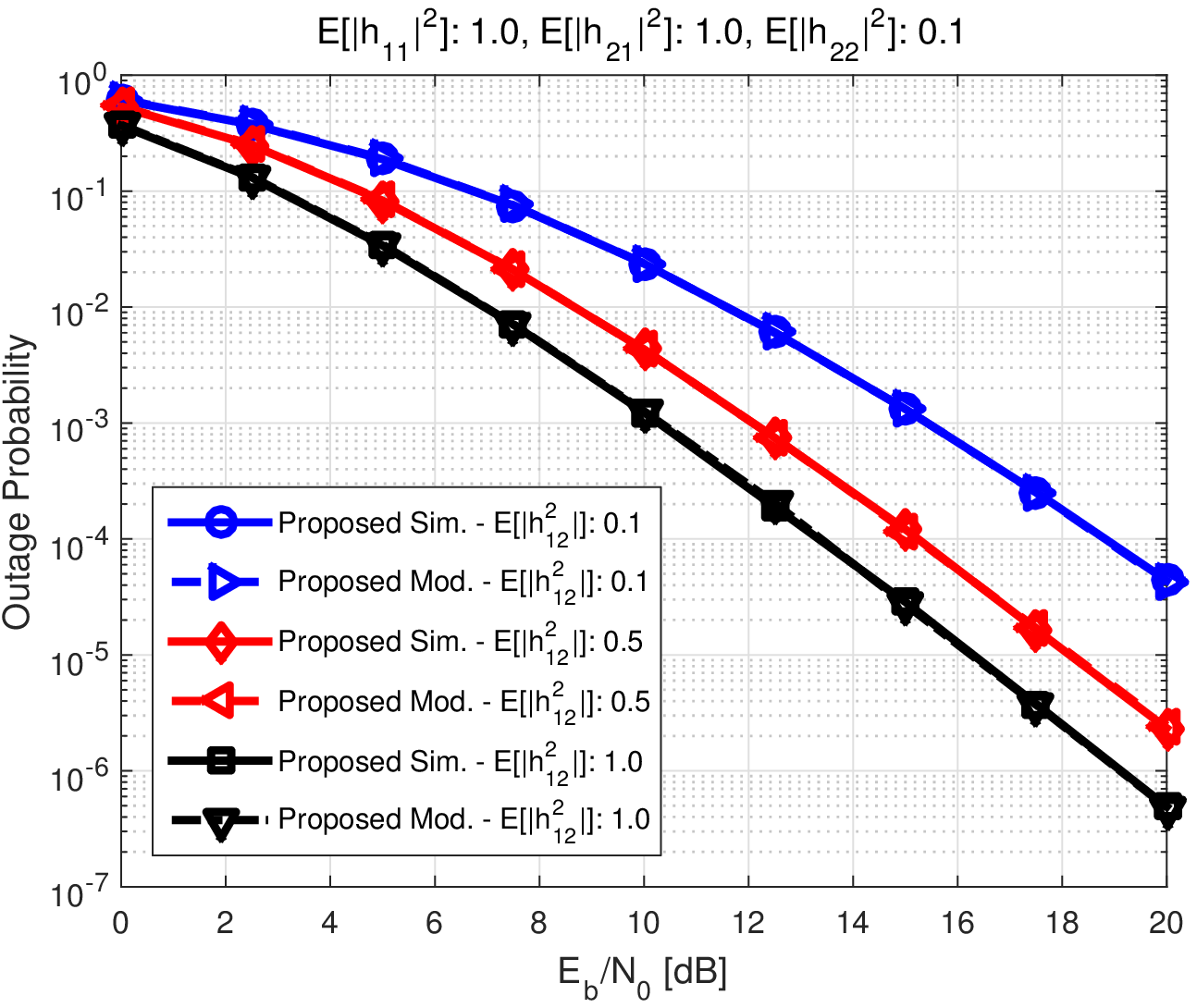}\label{fig:results4}}\hspace{-0.1cm}
\subfigure[$R_1$=$R_2$=1 bit/symbol]{\includegraphics[keepaspectratio,width = 0.32495\linewidth]{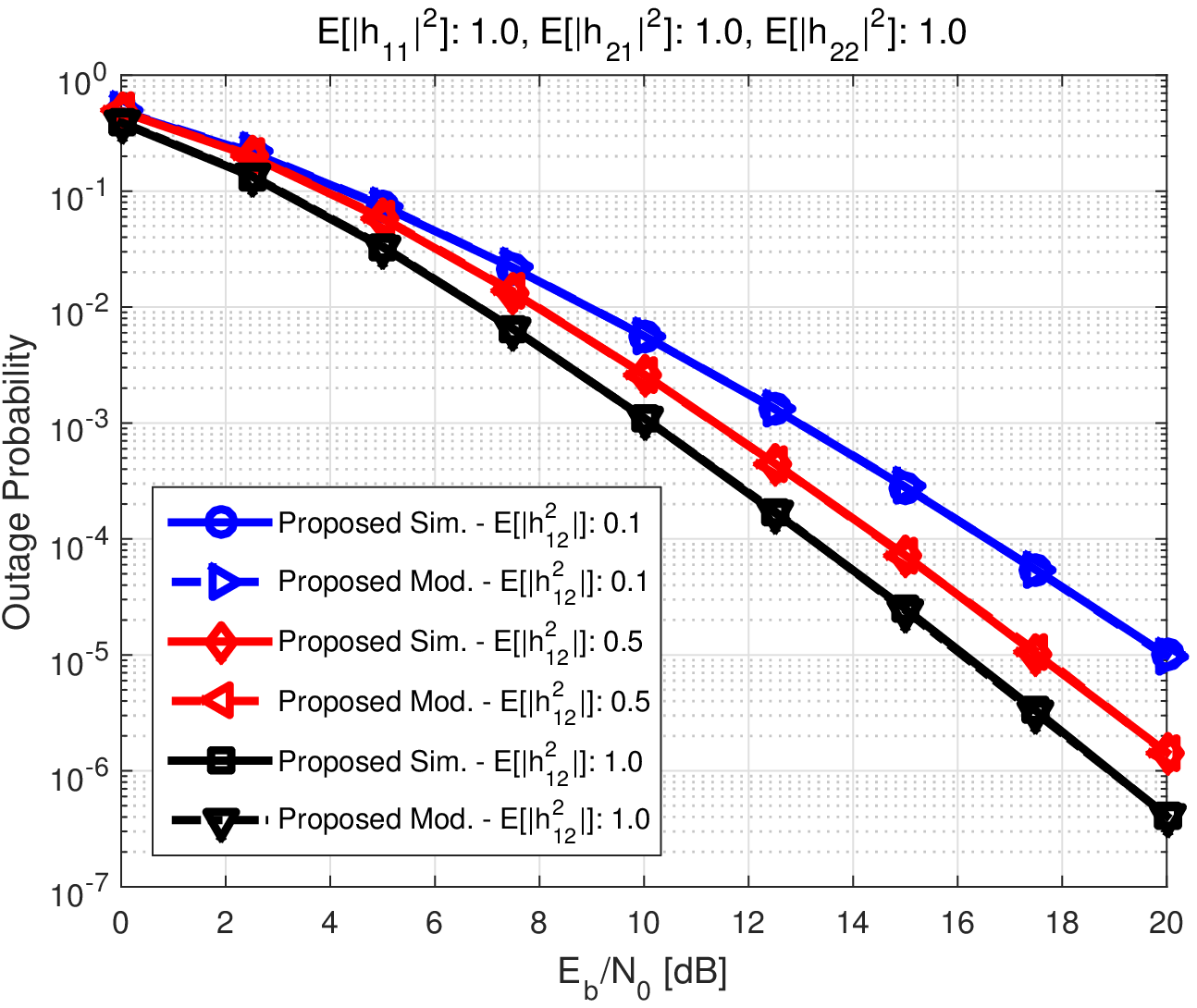}\label{fig:results3}}\hspace{-0.25cm}
\subfigure[Different MCSs.]{ \includegraphics[keepaspectratio,width = 0.32495\linewidth]{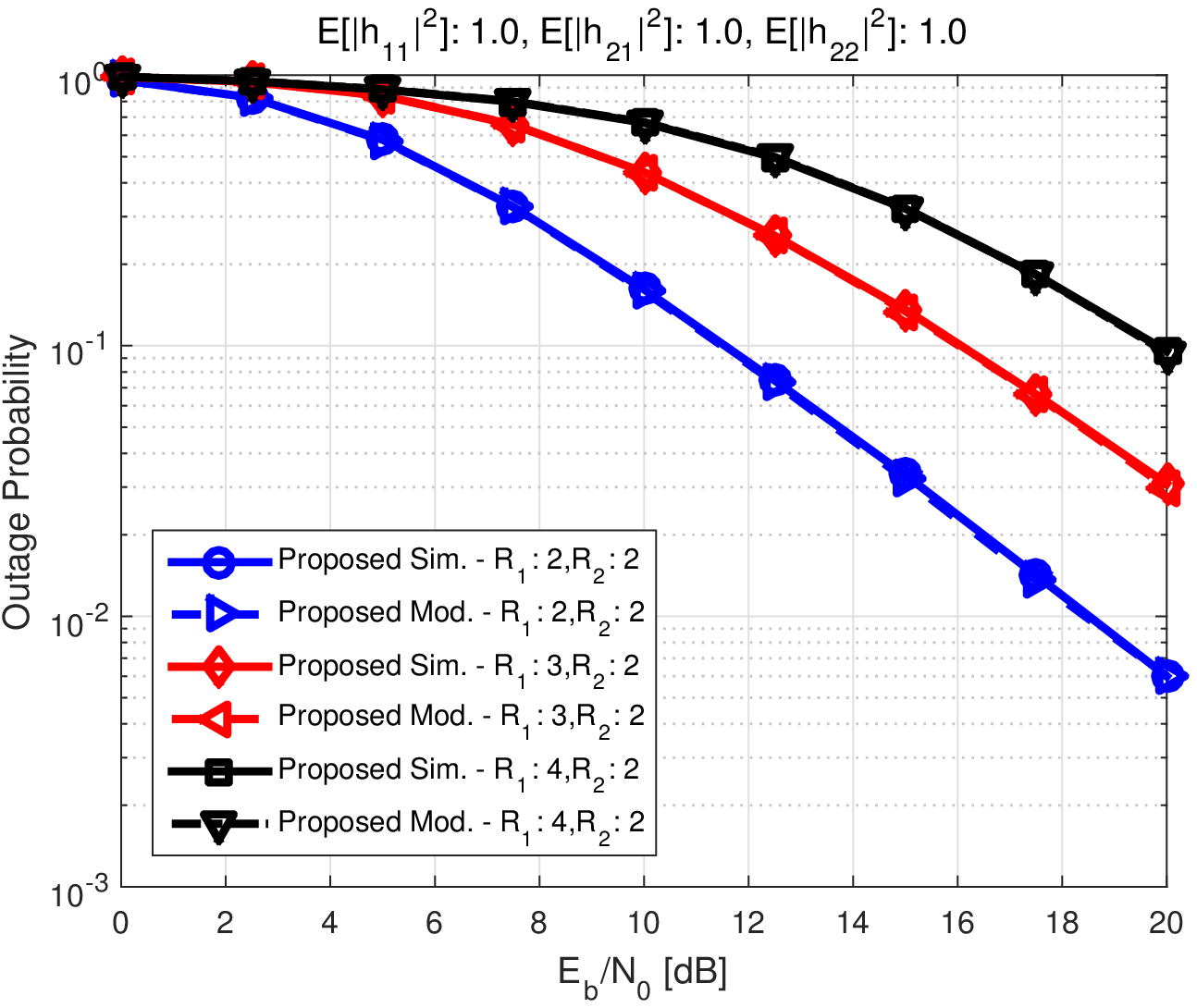}\label{fig:results13}}
\caption{Simulation and numerical results for the outage probability of the proposed protocol.}
\end{figure*}

For asymmetric links in Fig.~\ref{fig:results15} (i.e., $\Exp [|h_{1,2}|^2]$=0.1, $\Exp [|h_{2,1}|^2]$=0.1) the performance benefits of our scheme are even more evident over the complete SNR range. The benefits are because of the improved SIC decoding performance when the links are asymmetric. It is also important to see that in this case our scheme outperforms the DSTC/DF1 protocol simply because in that protocol the relay executes a 2-hop communication and only forwards when it  successfully decodes. Also note that CFNC matches the performance of DSTC/DF due to the lower number of network coding opportunities (since $\Exp [|h_{1,2}|^2]$=0.1 a small number of blocks from the second source are decoded in the same relay).

\textbf{Spectral efficiency comparisons.} Now we compare the SE of our protocol with that of a subset of the previous protocols again for $N_\text{R}$=$N_\text{RU}$=2. Besides CFNC, we also present results for DSTC/DF1. Transmitting with orthogonal DSTC protocols is suboptimal in every case as we already observed from the outage results and so we do not include them here. For the case of high inter-user interference and for $R_1$=$R_2$=2 we notice in Fig.~\ref{fig:results14_tput} that even though there was slightly worse outage performance of our proposed protocol when compared to CFNC after 20 dB, this is not translated to SE loss. The reason is that the outage probability is already very low for our protocol. Even more importantly, in this SNR regime the higher MCS with $R_1$=$R_2$=3 bit/symbol (that can be used by an adaptive protocol) leads to better performance of our protocol over CFNC even beyond 20 dB. This is important for obtaining gains with our protocol in every channel gain scenario.

Interesting results are also obtained when we compare the SE for low inter-user interference (i.e., $\Exp [|h_{1,2}|^2]$=0.1, $\Exp [|h_{2,1}|^2]$=0.1) in Fig.~\ref{fig:results15_tput}. We notice that our proposed protocol can achieve a spectral efficiency of 1.2 bits/symbol for a transmit SNR of 11 dB while CFNC is considerably lower. This particular average SNR is important since it is the borderline case where a higher order MCS is needed for better performance. The gains are even more pronounced for the higher order MCS of $R_1$=$R_2$=3 bit/symbol. Now DSTC/DF1~\cite{yao12} achieves lower gain simply because there is only one good link from the source to the destination while for the link from the second source to the destination it is $\Exp [|h_{2,d}|^2]$=0.1 in accordance with our setup in this paragraph. We can also see that we can obtain the same spectral efficiency in the high SNR regime also for our scheme. This result verifies the DMT analysis that predicts multiplexing gain of $\frac{2}{3}$ in the high SNR. It is also important to see that our scheme is better than the performance of DSTC/DF1~\cite{yao12} even with the disadvantage of two hops since the later protocol uses only DF. This illustrates again the important benefits of our protocol.

\begin{figure}[t]
\centering
 \subfigure[Diversity gains with 3 and 4 used relays simultaneously] {  \includegraphics[keepaspectratio,width = 0.4795\linewidth]{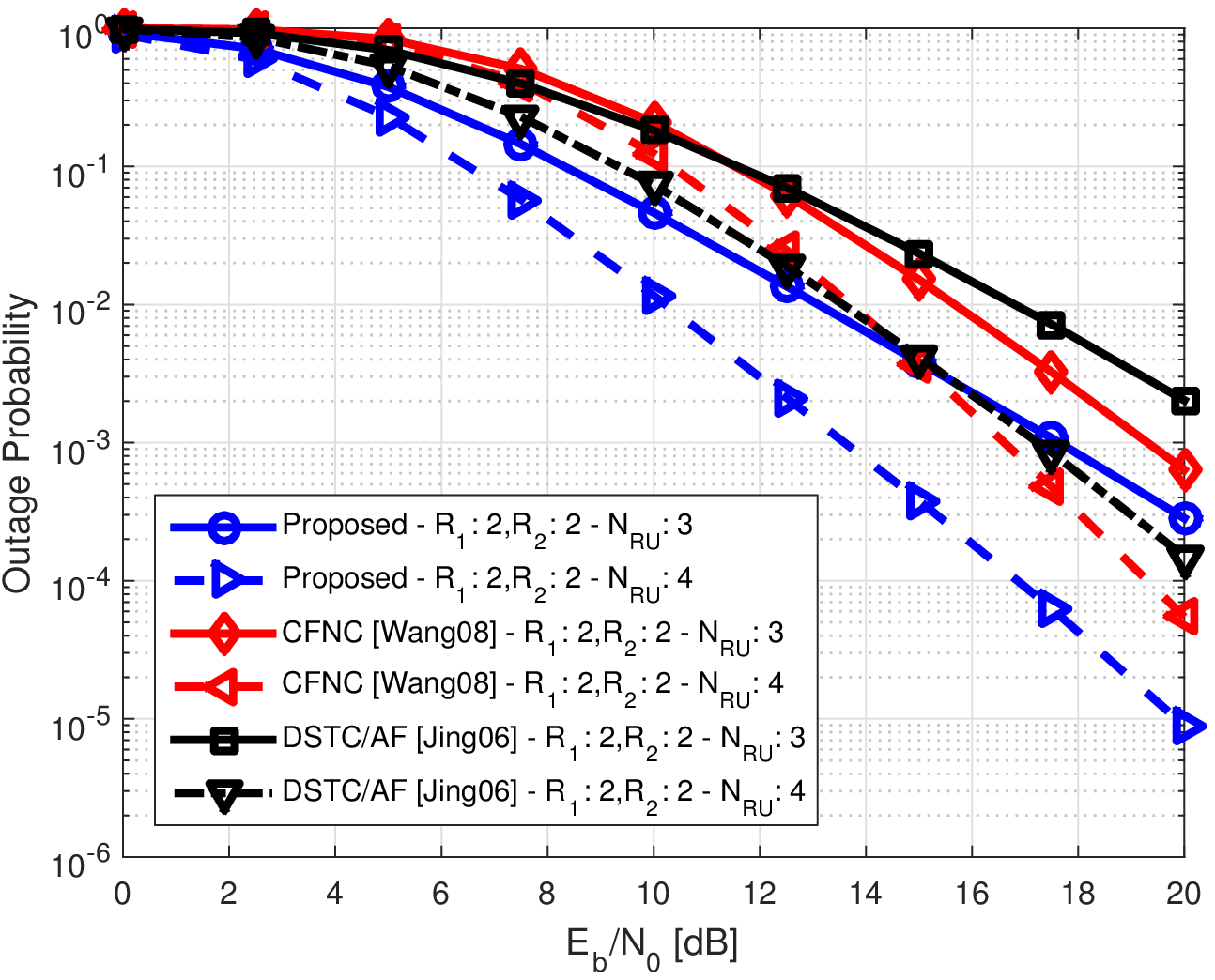}\label{fig:test_tput1}} \hspace{-0.3cm}
 \subfigure[The relay pre-selection algorithm for different number of potential relays] {
  \includegraphics[keepaspectratio,width = 0.4795\linewidth]{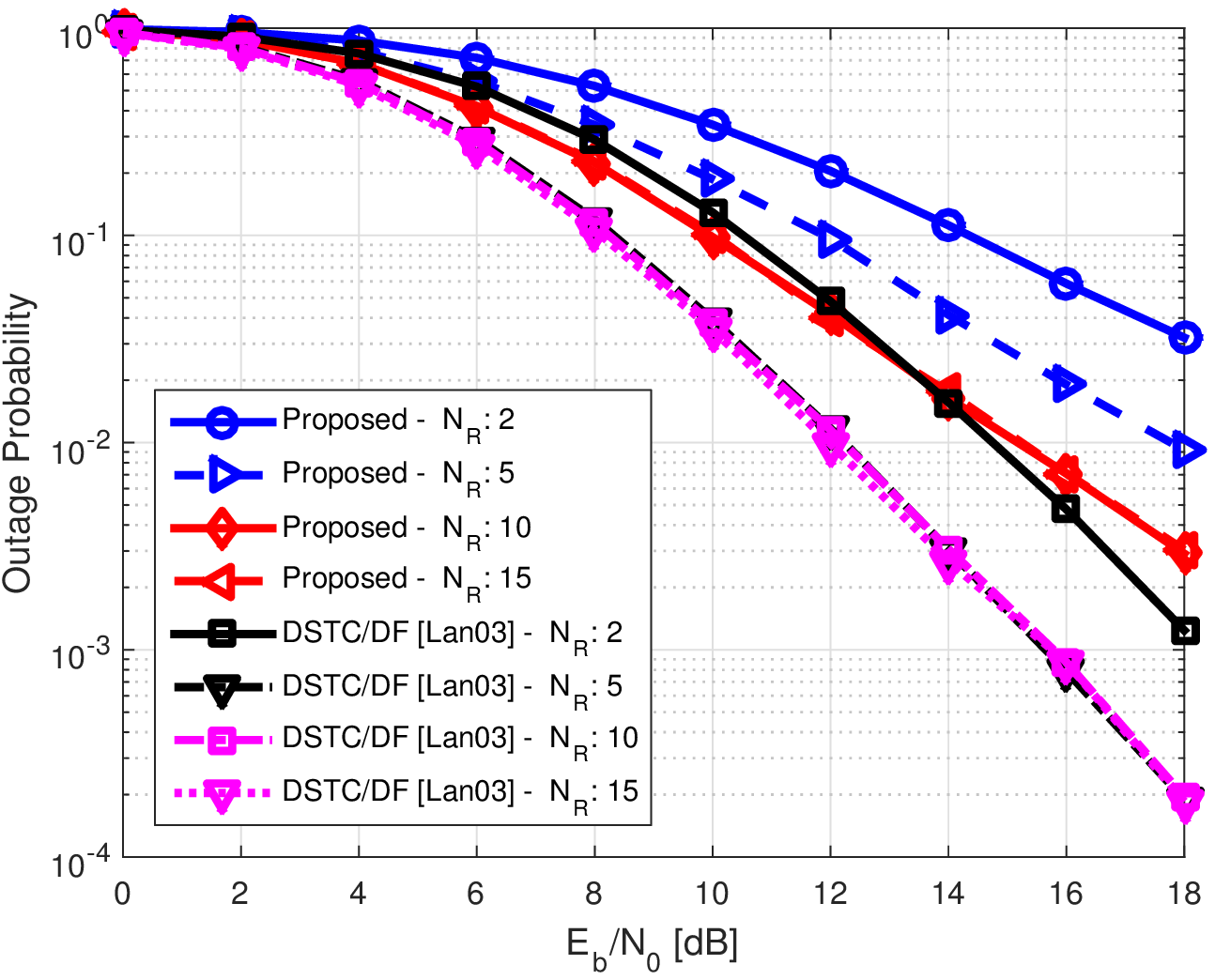}\label{fig:test_out17}
  }
\caption{Outage probability for multi-relay networks.}
\end{figure}

\textbf{Model validation.} We present numerical and simulation results for symmetric rate requirements of $R_1$=$R_2$=1 bit/symbol, $N_\text{R}$=$N_\text{RU}$=2, and different channel conditions for the compound MAC in first hop in Fig.~\ref{fig:results4}, and Fig.~\ref{fig:results3}. The parameters of the second hop are $\Exp [|f_{1}|^2]$=$\Exp [|f_{2}|^2]$=1.0. In Fig.~\ref{fig:results4} we set $\Exp [|h_{2,2}|^2]$=0.1. It is important to note that the performance of our protocol is dictated by the average channel gain from the source S1 to the second relay ($\Exp [|h_{1,2}|^2]$). This is because when $\Exp [|h_{1,2}|^2]$=1.0 our protocol can decode a strong signal from S1 that arrives at this second relay and so it can apply a STC for extracting a diversity gain of 2. The same is true also in Fig.~\ref{fig:results3} where we observe that the performance remains relatively unaffected even when the source S2 interferes significantly at the second relay ($\Exp [|h_{2,2}|^2] $=1.0 in this figure). Again the reason is that SIC behaves very well since a stronger signal from S2 towards RS2 ($\Exp [|h_{2,2}|^2]$ =1.0) can be cancelled more effectively. Our numerical results for this case follow closely the simulation. Numerical and simulation results agree very well because our analysis also considers the impact of SIC error propagation. Even though not clearly visible, we observed that the under-estimation of the simulation in the low SNR regime is because a fixed-length LDPC code was used. The results for different MCSs are shown in Fig.~\ref{fig:results13}, and they illustrate the validity of our model with respect to the different and asymmetric MCSs.

\textbf{Multi-relay diversity.} Multi-relay diversity is examined in Fig.~\ref{fig:test_tput1} for fully symmetric links. Now we evaluate a number of used relays equal to $N_\text{R}$=$N_\text{RU}$=3 and $N_\text{R}$=$N_\text{RU}$=4. A very important observation is that for higher $N_\text{RU}$ the bordeline case where CFNC outperforms our proposed protocol moves in the higher SNR regime unlike the case of $N_\text{RU}$=2 where this threshold was at 20 dB. Also both MU protocols outperform the classic orthogonal DSTC/AF that is known to achieve also full diversity.

\textbf{Multi-relay networks and pre-selection.} In the next scenario we have a network where $N_\text{R}$ relays are present but only $N_\text{RU}$ are pre-selected statically for simultaneous use. Here $R_1$=$R_2$=2 bit/symbol. The relays are spread randomly and uniformly in an area between the sources and the destination. In this case the distance between two nodes $i,j$ $\text{dist}(i,j)$, is normalized between 0 and 1, while we introduce a path loss model for the average channel gains, i.e., $\Exp[|h_{i,j}|^2]$=$1/ \text{dist}(i,j)^2$. In Fig.~\ref{fig:test_out17} the performance of the proposed system increases for high $N_\text{R}$ and saturates for $N_\text{R}$=15, 20. Even though there is no diversity gain from the presence of additional relays, i.e., higher $N_\text{R}$, since $N_\text{RU}$ is fixed to 2, better performance is achieved overall. Note that a classic orthogonal scheme that similarly uses $N_\text{RU}$=2 relays, cannot exploit the presence of more than 5 (three curves coincide).

\section{Conclusions}
\label{section:conclusions}
In this paper we introduced a novel cooperative protocol for a two-source multi-relay network that is based on opportunistic SIC and AF/DF relay operations. Analytical and simulation performance results show that significant diversity and multiplexing benefits can be observed over orthogonal DSTC protocols, and also multi-user protocols. We also introduced the static relay pre-selection concept enabled by our outage model that allows the exploitation of nodes as relays in dense networks with no practical overhead. Future work will investigate first the presence of more sources. Other avenues for future work include the optimization of the AF/DF functions, the study of the implications on the MAC protocol design, and the use of our idea in specific LTE and small-cell network topologies.

\section*{Acknowledgment}
The author would like to acknowledge the anonymous
reviewers for their detailed comments that helped improve
this manuscript considerably.



\end{document}